\documentclass[12pt]{article}
\usepackage{epsfig}
\usepackage{amssymb}
\usepackage{amsmath}
\usepackage{amsfonts}
\usepackage{graphicx}
\usepackage{mathrsfs}
\usepackage[dvips]{color}
\usepackage{multirow}
\usepackage{latexsym}

\newcommand{\R}{\mathbb{R}}
\newcommand{\C}{\mathbb{C}}
\newcommand{\Z}{\mathbb{Z}}

\newcommand{\ff}{\mathfrak{f}}
\newcommand{\fg}{\mathfrak{g}}

\newcommand{\fu}{\mathfrak{u}}

\newcommand{\fH}{\mathfrak{H}}

\newcommand{\fO}{\mathfrak{O}}

\newcommand{\fU}{\mathfrak{U}}

\newcommand{\fX}{\mathfrak{X}}

\newcommand{\bs}{{\mathbf{s}}}

\newcommand{\bH}{\mathbf{H}}

\newcommand{\bS}{\mathbf{S}}

\newcommand{\bU}{\mathbf{U}}

\newcommand{\bsigma}{{\boldsymbol{\sigma}}}

\newcommand{\cA}{{\mathcal{A}}}

\newcommand{\cH}{\mathcal{H}}
\newcommand{\cE}{\mathcal{E}}
\newcommand{\cF}{\mathcal{F}}
\newcommand{\cG}{\mathcal{G}}

\newcommand{\cO}{\mathcal{O}}
\newcommand{\cP}{\mathcal{P}}

\newcommand{\cS}{\mathcal{S}}
\newcommand{\cT}{\mathcal{T}}
\newcommand{\cU}{\mathcal{U}}

\newcommand{\cV}{\mathcal{V}}

\newcommand{\cX}{\mathcal{X}}

\newcommand{\be}{\begin{equation}}
\newcommand{\ee}{\end{equation}}
\newcommand{\bea}{\begin{eqnarray}}
\newcommand{\eea}{\end{eqnarray}}
\newcommand{\nn}{\nonumber}
\newcommand{\kt}{\rangle}
\newcommand{\br}{\langle}

\newcommand{\ed}{\end{document}}

\newcommand{\bi}{\begin{itemize}}
\newcommand{\ei}{\end{itemize}}

\newcommand{\emp}{\varnothing}
\newcommand{\bce}{\begin{center}}
\newcommand{\ece}{\end{center}}

\newcommand{\sD}{\mathscr{D}}
\newcommand{\sE}{\mathscr{E}}

\newcommand{\sH}{\mathscr{H}}

\newcommand{\sP}{\mathscr{P}}

\newcommand{\sT}{\mathscr{T}}
\newcommand{\sU}{\mathscr{U}}
\newcommand{\RE}{{\rm Re}}
\newcommand{\IM}{{\rm Im}}

\newcommand{\bigeta}{{\mbox{\large$\eta$}}}
\newcommand{\bigrho}{{\mbox{\large$\rho$}}}

\newcommand{\biggg}{{\mbox{\large$g$}}}
\newcommand{\bigfg}{{\mbox{\large$\fg$}}}
\newcommand{\bigh}{{\mbox{\large$h$}}}
\newcommand{\bigo}{{\mbox{\large$o$}}}

\newcommand{\bigu}{{\mbox{\large$u$}}}
\newcommand{\bigfu}{{\mbox{\Large$\fu$}}}

\newcommand{\smN}{{\mbox{\small $N$}}}
\newcommand{\smR}{{\mbox{\small $R$}}}

\newcommand{\ssA}{{\mbox{\scriptsize $A$}}}
\newcommand{\ssE}{{\mbox{\scriptsize $\sE$}}}
\newcommand{\scE}{{\mbox{\scriptsize $\cE$}}}
\newcommand{\ssN}{{\mbox{\scriptsize $N$}}}
\newcommand{\ssR}{{\mbox{\scriptsize $R$}}}

\makeatletter
\newcommand\xleftrightarrow[2][]{%
  \ext@arrow 9999{\longleftrightarrowfill@}{#1}{#2}}
\newcommand\longleftrightarrowfill@{%
  \arrowfill@\leftarrow\relbar\rightarrow}
\makeatother

\begin{document}

\title{Time-Dependent Pseudo-Hermitian Hamiltonians and a Hidden Geometric Aspect of Quantum Mechanics}

\author{Ali~Mostafazadeh\thanks{E-mail address: amostafazadeh@ku.edu.tr}~
\\ Departments of Mathematics 
and Physics, 
Ko\c{c} University,\\ 34450 Sar{\i}yer,
Istanbul, Turkey}

\date{ }
\maketitle

\begin{abstract}
A non-Hermitian operator $H$ defined in a Hilbert space with inner
product $\br\cdot|\cdot\kt$ may serve as the Hamiltonian for a
unitary quantum system, if it is $\bigeta$-pseudo-Hermitian for a
metric operator (positive-definite automorphism) $\bigeta$. The
latter defines the inner product $\br\cdot|\bigeta\cdot\kt$ of the
physical Hilbert space $\sH_{\eta}$ of the system. For situations
where some of the eigenstates of $H$ depend on time, $\bigeta$
becomes time-dependent. Therefore the system has a non-stationary
Hilbert space. Such quantum systems, which are also encountered in
the study of quantum mechanics in cosmological backgrounds, suffer
from a conflict between the unitarity of time evolution and the
unobservability of the Hamiltonian. Their proper treatment requires
a geometric framework which clarifies the notion of the energy
observable and leads to a geometric extension of quantum mechanics (GEQM). We provide a general introduction to the subject, review some of the recent developments, offer a straightforward description of the Heisenberg-picture formulation of the dynamics for quantum systems having a time-dependent Hilbert space, and outline the Heisenberg-picture formulation of dynamics in GEQM.



\end{abstract}

\section{Introduction}

The fact that a non-Hermitian operator can have a real spectrum is by no means unusual or surprising. For example, consider the operator $H:\C^2\to\C^2$ that is represented in the standard basis of $\C^2$ by the matrix
    \[ \bH:=\varepsilon \left[
    \begin{array}{cc}
    0 & 1\\
    4 & 0\end{array}\right],\]
where $\varepsilon$ is a positive real parameter. It is easy to check that $\bH$ and consequently $H$ have a pair of real eigenvalues namely $\pm 2\varepsilon$. In particular, they are diagonalizable and have a real spectrum. But does this mean that we can identify $H$ with an observable or the Hamiltonian of a quantum system? The answer to this question cannot be given unless we specify the inner product we wish to use for computing the expectation values of the observables of the system. If we adopt the standard Euclidean inner product $\br\cdot|\cdot\kt$, the answer is No. To see this, we recall that by definition, $\br\xi|\zeta\kt:=\xi_1^*\zeta_1+\xi_2^*\zeta_2 $ where
$\xi:=(\xi_1,\xi_2)$ and $\zeta:=(\zeta_1,\zeta_2)$ are arbitrary elements of $\C^2$, use $\sH$ to label the Hilbert space obtained by endowing $\C^2$ with the Euclidean inner product, and calculate the expectation value of $H$ in the state determined by the state vector $\chi:=\frac{1}{\sqrt 2}(1,-i)$. This gives
    \be
    \frac{\br \chi| H\chi\kt}{\br\chi|\chi\kt}=\frac{3i\varepsilon}{2}.
    \label{exp-val-1}
    \ee
Because this quantity is purely imaginary, we cannot interpret it as the average value of measurement outcomes $\pm 2\varepsilon$ which are real. This disqualifies $H$ to represent an observable of a quantum system with Hilbert space $\sH$, if we are to respect the measurement (projection) axiom of quantum mechanics (QM) \cite{dirac-qm,von-neumann}.

The fact that $H$ can have a complex expectation value is a manifestation of a basic result of linear algebra \cite{axler,schechter} which says: ``{\em A linear operator is Hermitian\footnote{Throughout this article we distinguish between operators and their matrix representations, for the latter depends on the choice of a basis. In particular, following von Neumann \cite{von-neumann}, we use the term ``Hermitian operator'' to mean ``self-adjoint operator,'' i.e., $H$ satisfies $\br\cdot,H\cdot\kt=\br H\cdot,\cdot\kt$, where $\br\cdot,\cdot\kt$ denotes the inner product of the Hilbert space. For a more precise definition see \cite{review}.} if and only if all its expectation values are real.}'' Because the reality of the expectation values is an indispensable ingredient of the measurement axiom, the claim that observables of a quantum system need not be Hermitian is false.

Another better known problem arises, if we try to identify $H$ with the Hamiltonian of a quantum system with Hilbert space $\sH$, i.e., demand that it generates the dynamics of the system via the Schr\"odinger equation,   
    \be
    i\partial_t\psi(t)=H\psi(t).
    \label{sch-eq}
    \ee
According to this equation,
    \bea
    \partial_t\br\psi(t)|\psi(t)\kt&=&\br\partial_t\psi(t)|\psi(t)\kt+\br\psi(t)|\partial_t\psi(t)\kt\nn\\
    &=&2\,\RE\left[\br\psi(t)|\partial_t\psi(t)\kt\right]
    \nn\\
    &=&2\,\IM\left[\br\psi(t)|H\psi(t)\kt \right],
    \label{x1}
    \eea
where ``$\RE$'' and ``$\IM$'' denote the real and imaginary part of their argument. Because $H$ has non-real expectation values the right-hand side of this equation can be nonzero. For example, letting $\psi(0):=\chi$ and using (\ref{exp-val-1}) and (\ref{x1}), we find $\partial_t\br\psi(t)|\psi(t)\kt\big|_{t=0}=3\varepsilon\neq 0$. This shows that the norm of the evolving state does change in time. Hence, $H$ does not generate a unitary time evolution.

The apparent conflicts with the measurement and unitarity axioms were responsible for the unpopularity of non-Hermitian operators among physicists interested in basic aspects of QM. For many decades their application was confined to effective theories which did not respect all of the Dirac-von Neumann axioms of QM. This situation drastically changed in the early 2000's after it was realized that a certain class of non-Hermitian operators can actually be made Hermitian upon a redefinition of the inner product of the Hilbert space \cite{p1,p2,p3,bbj-2002,jmp-2003,jpa-2003,jpa-2004}. The operator $H$ we considered above is a particular example. Let $\bigeta:\C^2\to\C^2$ and $\br\zeta,\xi\kt_{\eta}:\C^2\times\C^2\to\C$
be defined by
    \bea
    &&\bigeta\,\xi=\bigeta(\xi_1,\xi_2):=(\xi_1,\frac{\xi_2}{4}),
    \label{eta1}\\
    &&\br\zeta,\xi\kt_{\eta}:=\br\zeta|\bigeta\,\xi\kt:=\zeta_1^*\xi_1+\frac{\zeta_2^*\xi_2}{4}.
    \eea
Then, $\br\cdot,\cdot\kt_{\eta}$ defines a genuine (positive-definite) inner product \cite{review} in $\C^2$, and for every nonzero element $\xi=(\xi_1,\xi_2)$ of $\C^2$, we have
    \[\frac{\br \xi,H\xi\kt_\eta}{\br\xi,\xi\kt_\eta}=\frac{8\,\varepsilon\,\RE(\xi_1^*\xi_2)}{4|\xi_1|^2+|\xi_2|^2}.\]
This calculation shows that the expectation values of $H$ computed using the inner product $\br\cdot,\cdot\kt_{\eta}$ are real. Therefore, if we view $H$ as a linear operator acting in the Hilbert space $\sH_\eta$ defined by endowing $\C^2$ with the inner product $\br\cdot,\cdot\kt_{\eta}$, then it becomes Hermitian, i.e.,
    \[\br\zeta,H\xi\kt_{\eta}=\br H\zeta,\xi\kt_{\eta}.\]
This in turn implies the unitarity of the dynamics generated by the Schr\"odinger equation (\ref{sch-eq}) in the Hilbert space $\sH_\eta$, i.e., for each pair, $\phi(t)$ and $\psi(t)$, of solutions of this equation,
    \be
    \partial_t\,\br\phi(t),\psi(t)\kt_\eta=0.
    \nn
    \ee

The operator $\bigeta$ given by (\ref{eta1}) is an example of a metric operator acting in the Hilbert space $\sH$. We use the term ``metric operator'' to mean a positive-definite authomorphism\footnote{An automorphism is a one-to-one linear operator mapping all of $\sH$ onto $\sH$.}. This property ensures $\br\cdot,\cdot\kt_\eta$ to be a genuine positive-definite inner product. The requirement that $H$ is a Hermitian operator acting in $\sH_\eta$ is equivalent to demanding that it acts in $\sH$ as an $\bigeta$-pseudo-Hermitian operator, i.e.,
    \be
    H^\dagger=\bigeta\,H\,\bigeta^{-1},
    \label{ph}
    \ee
where $H^\dagger$ is the adjoint of $H$ viewed as an operator acting in $\sH$. The latter is defined by the condition: $\br\zeta|H^\dagger\xi\kt=\br H\zeta|\xi\kt$. We can also view $H$ as an operator acting in $\sH_\eta$ and introduce its adjoint $H^\sharp$ through the requirement: $\br\zeta,H^\sharp\xi\kt_\eta=\br H\zeta,\xi\kt_\eta$. It is not difficult to see that this is equivalent to
    \be
    H^\sharp:=\bigeta^{-1}H^\dagger\bigeta.
    \nn
    \ee
In light of this relation, we can identify (\ref{ph}) with $H^\sharp=H$, \cite{p1}. Therefore, $\bigeta$-pseudo-Hermitian operators acting in $\sH$ coincide with Hermitian operators acting in $\sH_\eta$. These constitute the observables of the quantum system determined by the Hilbert space-Hamiltonian operator pair $(\sH_\eta,H)$, \cite{jpa-2004}.

The notion of a pseudo-Hermitian operator as defined by (\ref{ph}) extends to situations where $\bigeta$ is a pseudo-metric operator, i.e., it is a Hermitian automorphism that needs not be positive-definite. In this more general setting and under the assumption that $H$ acts in a given Hilbert space $\sH$, has a discrete spectrum, and is diagonalizable (i.e., has a complete and bounded biorthonormal system \cite{review} formed out of its eigenvectors and those of its adjoint), one can prove that the following statements are equivalent \cite{p3}.
    \begin{itemize}
    \item[1)] $H$ is $\bigeta$-pseudo-Hermitian for a pseudo-metric operator $\bigeta$, i.e., it satisfies (\ref{ph}).
    \item[2)] The eigenvalues of $H$ are either real or come in complex-conjugate pairs.
    \item[3)] There is an antilinear operator $\cX$ that squares to identity and commutes with $H$.
    \end{itemize}
For situations where $H$ is expected to play the role the Hamiltonian of a quantum system the latter statement means that $\cX$ generates an antilinear symmetry of the system \cite{wigner}. This in turn clarifies the spectral consequences of $\cP\cT$-symmetry \cite{bender-1998,bender-jmp-1999,delabaere-2000,dorey}.

With the stronger requirement that $\bigeta$ be positive-definite one can establish the reality of the spectrum of $H$, its quasi-Hermiticity\footnote{A linear operator $H$ is called quasi-Hermitian \cite{SGH} if it is related to a Hermitian operator $h$ by a similarity transformation generated by a positive-definite automorphsim $\rho$, i.e., $H=\rho^{-1}h\,\rho$.}, and the exactness of the antilinear symmetry $\cX$. More precisely the following statements are equivalent \cite{p3,jpa-2003}.
    \begin{itemize}
    \item[1$'$)] $H$ is $\bigeta$-pseudo-Hermitian for a metric operator $\bigeta$.
    \item[2$'$)] $H$ acts as a Hermitian operator in $\sH_\eta$.
    \item[3$'$)] The eigenvalues of $H$ are real.
    \item[4$'$)] The operator $\bigh:=\bigrho^{-1}H\,\bigrho$ with
    $\bigrho:=\sqrt\bigeta$ acts as a Hermitian operator in $\sH$.\footnote{Here $\sqrt\bigeta$ stands for the positive square root of $\bigeta$.} In particular as an operator acting in $\sH$, $H$ is quasi-Hermitian.
    \item[5$'$)] There is an antilinear operator $\cX$ that squares to identity, and there is a complete set of common eigenvectors of $H$ and $\cX$.
     \end{itemize}

Suppose that the statement 1$'$ holds, so that
$H:\sH_\eta\to\sH_\eta$ is Hermitian. Then we can identify
$\sH_\eta$ and $H$ with the Hilbert space and Hamiltonian of a
quantum system $\cS$. Being a Hermitian operator acting in
$\sH_\eta$, $H$  determines an observable of $\cS$. Furthermore,
because Hermitian operators have real expectation values, a
calculation similar to the one leading to (\ref{x1}) implies that
$H$ generates unitary evolutions. Hence $\cS$ is a unitary quantum
system. An alternative way of arriving at this conclusion is to note
that the Hilbert space-Hamiltonian operator pair $(\sH,\bigh)$ also
describes the same quantum system $\cS$. Indeed, because
    \[\br\bigrho\,\zeta|\bigrho\,\xi\kt=\br\zeta|\bigrho^2\xi\kt=
    \br\zeta|\bigeta\,\xi\kt=\br\zeta,\xi\kt_\eta,\]
$\bigrho$ defines a unitary operator\footnote{In standard texts on quantum mechanics, a unitary operator $\sU$ is defined as a linear mapping that maps a given Hilbert space onto the same Hilbert space and preserves the inner product of vectors. Here we use the standard generalization of this notion to the case that the operator $\sU$ maps a Hilbert space $\sH_1$ with inner product $\br\cdot,\cdot\kt_1$ onto another Hilbert space $\sH_2$ with inner product $\br\cdot,\cdot\kt_2$. If domain of $\sU$ is $\sH_1$, and for all $\psi_1,\phi_1\in\sH_1$ we have $\br\psi_1,\phi_1\kt_1=\br\sU\psi_1,\sU\psi_2\kt_2$, we say that $\sU$ is a unitary operator.}
mapping $\sH_\eta$ to $\sH$,
\cite{jpa-2003}. This in turn implies that if $\psi\in\sH_\eta$ and
$O:\sH_\eta\to\sH_\eta$ respectively describe a state and an
observable of $\cS$, $\Psi:=\bigrho\,\psi\in\sH$ and
$\bigo:=\bigrho\,O\bigrho^{-1}:\sH\to\sH$ describe the same state
and observable of $\cS$. This is simply because both choices lead to
the same expectation values;
    \[\frac{\br\psi,O\psi\kt_\eta}{\br\psi,\psi\kt_\eta}=
    \frac{\br\psi|\bigeta O\psi\kt}{\br\psi|\bigeta\psi\kt}=
    \frac{\br\psi|\bigrho^2 O\psi\kt}{\br\psi|\bigrho^2\psi\kt}=
    \frac{\br\bigrho\psi|\bigo\bigrho \psi\kt}{\br\bigrho\psi|\bigrho\psi\kt}=
    \frac{\br\Psi|\bigo\,\Psi\kt}{\br\Psi|\Psi\kt}.\]
To sum up,  $(\sH_\eta,H)$ and $(\sH,\bigh)$ provide different
mathematical  representations of the same quantum system
\cite{jpa-2004}. In particular, we can use either of them to
determine the physical properties of this system.

The initial work on pseudo-Hermitian operators \cite{p1,p2,p3} was
motivated by the need for a careful evaluation of the prospects of
$\cP\cT$-symmetric QM \cite{bender-jmp-1999} and the possible
relevance of these operators to certain constructions arising in the
two-component formulation of the mini-superspace Wheeler-DeWitt
equation \cite{jmp-1998}.

The results reported in Refs.~\cite{jpa-2003,jpa-2004,cjp-2004}
showed that indeed certain $\cP\cT$-symmetric Hamiltonian operators
are capable of defining unitary quantum systems, but these systems
also admitted a description in terms of Hermitian Hamiltonian
operators. Therefore, the use of $\cP\cT$-symmetric
(and more generally pseudo-Hermitian) Hamiltonians do
not actually yield a generalization of QM. It rather gives rise to
previously unexplored equivalent representations of quantum
mechanics \cite{review}.

An important by-product of the study of pseudo-Hermitian operators
was the introduction of new technologies for the construction of
inner products \cite{p1,p2,jmp-2006}. For certain physically interesting
quantum cosmological models, these could be employed for the purpose
of endowing the solution space of the Wheeler-DeWitt equation with
the structure of a genuine Hilbert space \cite{cqg-2003,ap-2004}.
This meant solving the infamous Hilbert-space problem
\cite{kuchar} for these models. The same approach allowed for a
complete and consistent formulation of QM of a first-quantized free
Klein-Gordon field \cite{ijmpa-2006,ap-2006}, a Proca field
\cite{jmp-2009}, and more recently a free photon \cite{jmp-2017,Hawton-2019}.

Quantum cosmological applications of pseudo-Hermitian operators
require dealing with time-dependent metric operators
\cite{cqg-2003,ap-2004}. For a quantum system represented by the
Hilbert space-Hamiltonian operator pair $(\sH_\eta,H)$, the proof of
the unitarity of time-evolution encounters a major difficulty
whenever $\bigeta$ depends on time. More precisely, the requirement
of unitarity of dynamics conflicts with the
$\bigeta$-pseudo-Hermiticity and hence observability of the
Hamiltonian. Since its announcement \cite{plb-2007} in 2007, there
have appeared different proposals for resolving this conflict in the
literature
\cite{znojil-2008,znojil-2013,gong-2013,znojil-2015,fring-2016,znojil-2017}.
A careful assessment of the geometric aspects of this problem has
recently led to a comprehensive resolution that not only clarifies
the role of the energy operator for quantum systems having a dynamical
Hilbert space, but also paves the way towards a geometric
extension of quantum mechanics (GEQM) \cite{prd-2018}. In the
present article, we provide a brief review of these developments,
discuss their conceptual implications, and outline a
Heisenberg-picture formulation of the dynamics for systems with a
time-dependent state space and systems considered in the framework of GEQM.

\section{Time-dependent pseudo-Hermiticity}

Consider a quantum system $\cS$ represented by the Hilbert
space-Hamiltonian operator pair $(\sH_\eta,H)$, where $\bigeta$ is a
time-dependent metric operator,  and let $\psi$ and $\phi$ be
arbitrary solutions of the Schr\"odinger equation (\ref{sch-eq}).
Then,
    \bea
    \partial_t\br\phi|\psi\kt_{\eta}&=&\partial_t\br\phi|\bigeta\,\psi\kt\nn\\
    &=&\br\dot\phi|\bigeta\,\psi\kt+\br\phi|\bigeta\,\dot\psi\kt+\br\phi|\dot{\bigeta}\,\psi\kt\nn\\
    &=&\br-i H\phi|\bigeta\,\psi\kt+\br\phi|-i\bigeta\,H\psi\kt+\br\phi|\dot{\bigeta}\,\psi\kt\nn\\
    &=&i\br\phi|(H^\dagger\bigeta-\bigeta\,H-i\dot\bigeta)\psi\kt\nn\\
    &=&i\br\phi,(H^\sharp-H-i\bigeta^{-1}\dot\bigeta)\psi\kt_\eta,
    \label{x2}
    \eea
where an overdot labels a time derivative. In order for $H$ to
generate a unitary evolution, the right-hand side of (\ref{x2})
must vanish for every choice of the solutions $\phi$ and $\psi$.
This happens if and only if
    \be
    H^\sharp=H+i\bigeta^{-1}\dot\bigeta.
    \label{xx0}
    \ee
Because $\bigeta$ is time-dependent and $\bigeta^{-1}$ is
invertible,  this equation implies, $H^\sharp\neq H$, i.e., $H$ is not a
Hermitian operator acting in $\sH_\eta$. Therefore, if $H$ generates
a unitary dynamics, it does not correspond to an observable of the
quantum system $\cS$! This is the content of the conflict between
the unitarity of the time evolution generated by the Schr\"odinger
equation (\ref{sch-eq}) in $\sH_\eta$ and the observability of the
Hamiltonian $H$, \cite{plb-2007}.

The initial work  on the construction of the most general metric
operator $\bigeta$ for a diagonalizable Hamiltonian $H$ with a real
and discrete spectrum \cite{p1,p2} revealed the following spectral
expansion of $\bigeta$.
    \be
    \bigeta=\sum_{n} |\phi_{n}\kt\br\phi_{n}|,
    \label{x3}
    \ee
where $\phi_{n}$  are eigenvectors of $H^\dagger$ that constitute a
(Riesz) basis of the Hilbert space \cite{review}, and for every
$\zeta\in\sH$, the symbol $|\zeta\kt\br\zeta|$ stands for the linear
operator that maps state vectors $\xi$ to $\br\zeta|\xi\kt\zeta$. A
simple consequence of (\ref{x3}) is that unless $H$ and therefore
$H^\dagger$ have a complete set of time-independent eigenvectors,
every metric operator $\bigeta$ that renders $H$ pseudo-Hermitian is
necessarily time-dependent. This underlines the significance of
addressing the conflict between unitarity and the observability of
generic time-dependent Hamiltonians.

There are essentially three different ways of dealing with this
conflict:
    \begin{itemize}
    \item[ i)] Modifying the Schr\"odinger equation to avoid this conflict.
    \item[ii)] Upholding unitarity at the expense of unobservability of the Hamiltonian.
    \item[iii)] Abandoning the requirement of unitarity in favor of the observability of the Hamiltonian.
    \end{itemize}
To the best of our knowledge option iii was never considered as viable, while there appeared a number of publications \cite{znojil-2008,znojil-2013,gong-2013,znojil-2015,fring-2016,znojil-2017} advocating options i or ii. The developments reported in these publications rest on the following premises:
    \begin{itemize}
    \item[a)] There is a representation of $\cS$ defined by the Hilbert space $\sH$ and a generally time-dependent Hermitian Hamiltonian operator $\bigh$ acting in $\sH$. This operator generates the dynamics of the state vectors in $\sH$ via the standard Schr\"odinger equation,
     \be
     i\partial_t\Psi(t)=\bigh(t)\Psi(t),
     \label{sch-eq-h}
     \ee
   and identifies an observable of the system which is customarily called the {\em energy observable}.
    \item[b)] Given a possibly time-dependent metric operator $\bigeta$, we can represent $\cS$ using the Hilbert space $\sH_\eta$ and an operator $H$ that generates time evolutions in $\sH_\eta$, such that the unitary transformation $\bigrho^{-1}:\sH\to\sH_\eta$ maps the solutions of the Schr\"odinger equation (\ref{sch-eq-h}) defined by $\bigh$ to those of the Schr\"odinger equation (\ref{sch-eq}) defined by $H$. It is easy to show that this condition is equivalent to the requirement:
        \be
        H=\bigrho^{-1} \bigh \bigrho-i\bigrho^{-1}\dot\bigrho.
        \label{x4}
        \ee
    \item[c)] In the representation $(\sH_\eta,H)$, the observables of $\cS$, which are represented by Hermitian operators $O$ acting in $\sH_\eta$, are obtained from their representatives $\bigo$ in the representation $(\sH,\bigh)$ via $O=\bigrho^{-1} \bigo \bigrho$. In particular,  in the representation $(\sH_\eta,H)$, the energy observable is represented by
    \be
    H_\ssE:=\bigrho^{-1} \bigh \bigrho.
    \label{x05}
    \ee
    \end{itemize}

If we insist that the Hamiltonian  and the energy observable must
coincide in both of the representations, $(\sH,\bigh)$ and
$(\sH_\eta,H)$, we have no choice but to agree that, in
the representation $(\sH_\eta,H)$, the dynamical evolution of the
state vectors is determined by the modified Schr\"odinger equation
\cite{znojil-2008,gong-2013},
    \be
    i\sD_t\psi=H_\ssE\psi,
    \label{mod-sch-eq}
    \ee
where
    \be
    \sD_t:=\partial_t+\bigrho^{-1}\dot\bigrho.
    \label{covariant}
    \ee
This provides a resolution of the unitarity versus observability
conflict via a modification of the Schr\"odinger equation. Note,
however, that this approach stems from a particular choice
of terminology. We could simply refrain from using the term
``Hamiltonian'' for the ``energy operator,'' but instead take the
former to mean the ``generator of time evolutions'' determined by
the usual Schr\"odinger equation~(\ref{sch-eq}). We are then led to
the inevitable conclusion that the Hamiltonian is not an observable
unless $\bigrho$ and consequently $\bigeta$ are time-independent
\cite{fring-2016}. This is in line with the resolution ii of the
above-mentioned conflict.

\section{Dynamical inner products realizing unitarity}

In specific applications in quantum cosmology
\cite{cqg-2003,ap-2004}, the generator of time evolutions is the
only input of the problem, and the aim is to determine an
appropriate Hilbert space in which the time evolution is realized
via a one-parameter family of unitary operators. If one can identify
a Hilbert space $\sH$ in which the generator of time evolutions acts
as a linear operator with a real and discrete spectrum and there is
complete and bounded biorthonormal system \cite{review} consisting of the
eigenvectors of this operator and its adjoint, then there are metric
operators $\bigeta$ such that this operator is
$\bigeta$-pseudo-Hermitian. However, for cases where all the metric
operators $\bigeta$ with this property are time-dependent, we cannot
establish the unitarity of the time evolution by working in the
Hilbert space $\sH_\eta$. Ref.~\cite{cqg-2003} offers a solution for
this problem that involves finding metric operators $\bigeta$ that
achieve the unitarity of the time evolutions, not the
$\bigeta$-pseudo-Hermiticity of their generator.

Let $H(t)$ label the generator of time evolutions, and $U(t,t_0)$ be the corresponding
evolution operator for the initial time $t_0$, so that $i\partial_t
U(t,t_0)=H(t)U(t,t_0)$ and $U(t_0,t_0)=I$, where $I$ is the identity
operator acting in $\sH$. We can express the unitarity of dynamical
evolutions in $\sH_{\eta(t)}$ in the form
    \[\br\phi(t),\psi(t)\kt_{\eta(t)}=\br\phi(t_0),\psi(t_0)\kt_{\eta(t_0)}.\]
This relation implies that for every choice of initial state vectors
$\phi(t_0):=\phi_0$ and $\psi(t_0):=\psi_0$,
    \bea
    \br\phi_0|\bigeta(t_0)\psi_0\kt&=&\br\psi(t)|\bigeta(t)\psi(t)\kt\nn\\
    &=&\br U(t,t_0)\phi_0|\bigeta(t)U(t,t_0)\psi_0\kt\nn\\
    &=&\br\phi_0|U(t,t_0)^\dagger\bigeta(t)U(t,t_0)\psi_0\kt.\nn
    \eea
This is true for every $\phi_0,\psi_0\in\sH$ if and only if
    \be
    \bigeta(t)={U(t,t_0)^\dagger}^{-1}\bigeta_0\,U(t,t_0)^{-1},
    \label{x5}
    \ee
where $\bigeta_0:=\bigeta(t_0)$. Eq.~(\ref{x5}) determines the
metric operator $\bigeta(t)$ and consequently $\sH_{\eta(t)}$ up to
the choice of $\bigeta_0$. A suitable choice, which is however not
dictated by the details of the problem at hand, is to identify
$\bigeta_0$ with a metric operator so that $H(t_0)$ is
$\bigeta_0$-pseudo-Hermitian \cite{cqg-2003,ap-2004}. This in turn
implies that $H(t_0)$ is an observable of the system represented by
$(\sH_{\eta(t)},H(t))$ at time $t_0$, but for $t\neq t_0$ the same
does not generally apply to $H(t)$.\footnote{According to
(\ref{xx0}), the requirement of the  $\bigeta_0$-pseudo-Hermiticity
of $H(t_0)$ is equivalent to $\dot\bigeta(t_0)=0$.} Notice however
that there is a priori no reason to assume that $H(t_0)$ is
$\bigeta_0$-pseudo-Hermitian for some metric operator $\bigeta_0$.

An important observation regarding (\ref{x5}) is that it provides
the general solution of (\ref{xx0}) when we view the latter as an
equation for $\bigeta$. Using this equation, we can actually check
that
    \be
    \bigh(t):=\bigrho(t)H(t)\bigrho(t)^{-1}+i\dot\bigrho(t)\bigrho(t)^{-1}
    \label{xx-28}
    \ee
is a Hermitian operator acting in $\sH$. Furthermore, because it
satisfies (\ref{x4}), $\bigrho(t):=\sqrt{\bigeta(t)}$ maps the
solutions of the Schr\"odinger equation (\ref{sch-eq}) for the
Hamiltonian $H(t)$ to those of the Schr\"odinger equation
(\ref{sch-eq-h}) for the Hamiltonian $\bigh(t)$. By virtue of the
fact that $\bigrho(t):\sH_{\eta(t)}\to\sH$ is a unitary operator,
this shows that $(\sH,\bigh(t))$ and $(\sH_{\eta(t)},H(t))$
represent the same quantum system. A rather unexpected aspect of the
latter representation is that not only $H(t)$ fails to be
$\bigeta(t)$-pseudo-Hermitian, but indeed it may happen not to be a
pseudo-Hermitian operator at all, i.e., there may exist no metric operator
$\tilde\bigeta(t)$ such that $H(t)$ is
$\tilde\bigeta(t)$-pseudo-Hermitian.

As a simple example, consider the situation where $\sH$ is the Hilbert space
of square-integrable functions and
    \be
    H(t):=H_0(t)+\ff(t)\cP,
    \label{pert-sho}
    \ee
where $H_0:=P^2/2m+m\omega^2 X^2/2$ is the standard Hamiltonian for
a simple harmonic oscillator with mass $m$ and angular frequency
$\omega$, $X$ and $P$ are the standard position and momentum
operators acting in $\sH$, $\ff:\R\to\C$ is a piecewise continous
complex-valued function of time, and $\cP$ is the parity operator
defined by $(\cP\psi)(x):=\psi(-x)$.

Because $H_0$ and $\cP$ act in $\sH$ as commuting Hermitian
operators, the spectrum of $H(t)$ consists of the eigenvalues of the
form $\omega(n+1/2)\pm\ff(t)$, where $n$ is a nonnegative integer.
This shows that for the cases where $\ff(t)$ is neither real nor
imaginary, $H(t)$ is not pseudo-Hermitian. Yet we can compute its
evolution operator and use (\ref{x5}) to determine a metric operator
that makes the time evolution generated by $H(t)$ unitary. Setting
$\bigeta_0=I$, so that $\sH_{\eta(t_0)}=\sH$, we find
    \begin{align}
    &U(t,t_0)=U_0(t,t_0)e^{-i\cF(t)\cP},
    &&\bigeta(t)= e^{-2\,\IM[\cF(t)]\cP},
    &&\bigrho(t)= e^{-\,\IM[\cF(t)]\cP},
    \label{xx32}
    \end{align}
where $U_0(t,t_0):=\exp[-i(t-t_0)H_0]$ is the time-evolution
operator for the simple harmonic oscillator, and
$\cF(t):=\int_{t_0}^t\ff(t')dt'$. Substituting (\ref{pert-sho}) in
(\ref{xx-28}) and using the last relation in (\ref{xx32}), we have
    \[\bigh(t)=H_0+\RE[\ff(t)]\,\cP.\]
This shows that the quantum system represented by
$(\sH_{\eta(t)},H(t))$ also admits the representation
$(\sH,\bigh(t))$.

If we identify $\bigh(t)$ with the energy observable of the system
in the representation $(\sH,\bigh(t))$, then in view of (\ref{x05})
and (\ref{xx32}) the operator $H_\ssE(t)$ representing this
observable in $(\sH_{\eta(t)},H(t))$ coincides with $\bigh(t)$. This
is not generally true for other observables. For example, in the
representation $(\sH_{\eta(t)},H(t))$, the position and momentum
operators are given by \cite{jpa-2006a}:
    \begin{align}
    &x_\eta:=\bigrho(t)^{-1}X\bigrho(t)=e^{2\IM[\cF(t)]\cP}X,
    &p_\eta:=\bigrho(t)^{-1}P\bigrho(t)=e^{2\IM[\cF(t)]\cP}P.
    \nn
    \end{align}

If we insist on using the term ``Hamiltonian'' for the energy
operator $H_\ssE$ and demand that this operator generates the dynamics
via a first-order linear differential equation involving $H_\ssE$,
we are led to the modified Schr\"odinger equation (\ref{mod-sch-eq})
with $H_\ssE(t)=\bigh(t)$ and
    \[\sD_t:=\partial_t-\,\IM[\ff(t)]\,\cP.\]

\section{Heisenberg picture of dynamics}

The description of the dynamics of a quantum system in the
Heisenberg picture has many advantages. The study of the Heisenberg
picture for a unitary quantum system defined by a time-independent
pseudo-Hermitian Hamiltonian or a Hamiltonian acting in a
time-dependent Hilbert space has been considered in
Refs.~\cite{znojil-2015,Miao-2016}. In this section we provide our
approach for addressing this problem.

Consider the representation $(\sH,\bigh(t))$ of our generic quantum
system $\cS$ where observables are given by Hermitian operators
$\bigo(t):\sH\to\sH$, and the dynamics of state vectors is generated
by the Hermitian Hamiltonian operator $\bigh(t)$. In the Heisenberg
picture, the state vectors are stationary while the operators
corresponding to observables evolve in time according to
    \be
    \bigo(t_0)\longrightarrow\bigo^{\rm(H)}(t):=\bigu(t,t_0)^{-1}\bigo(t)\bigu(t,t_0).
    \label{HP1}
    \ee
Here $\bigu(t,t_0)$ is the time-evolution operator corresponding to the Hamiltonian $\bigh(t)$ and the initial time $t_0$, i.e., the operator satisfying
    \begin{align}
    &i\partial_t\bigu(t,t_0)=\bigh(t)\bigu(t,t_0),
    &&\bigu(t_0,t_0)=I.
    \label{HP2}
    \end{align}
If we differentiate both sides of (\ref{HP1}) and use (\ref{HP2}) to simplify the result, we obtain the Heisenberg equation of motion in the representation $(\sH,\bigh(t))$:
    \be
    i\partial_t \bigo^{\rm(H)}(t)=[\bigo^{\rm(H)}(t),\bigh^{\rm(H)}(t)]+i
    \bigu(t,t_0)^{-1}\dot\bigo(t)\bigu(t,t_0),
    \label{HP3}
    \ee
where $\bigh^{\rm(H)}(t):=\bigu(t,t_0)^{-1}\bigh(t)\bigu(t,t_0)$ is
the Heisenberg-picture Hamiltonian.

Next, we examine the Heisenberg equation in the representation
$(\sH_{\eta(t)},H(t))$. To derive this equation, we use the fact
that if an observable is given by the operator $\bigo(t)$ in the
representation $(\sH,\bigh(t))$ of the system $\cS$, then it is
given by
    \be
    O(t):=\bigrho(t)^{-1}\bigo(t)\bigrho(t),
    \label{HP4}
    \ee
in the representation $(\sH_{\eta(t)},H(t))$, \cite{jpa-2004}. We
also recall that the Heisenberg-picture operator corresponding to
(\ref{HP4}) has the form
    \be
    O^{\rm(H)}(t):=U(t,t_0)^{-1}O(t)U(t,t_0).
    \label{HP5}
    \ee
In particular,
    \be
    H^{\rm(H)}(t):=U(t,t_0)^{-1}H(t)U(t,t_0),
    \label{HP6}
    \ee
gives the expression for the Heisenberg-picture Hamiltonian in the representation $(\sH_{\eta(t)},H(t))$. Furthermore, because $\bigrho(t):\sH_{\eta(t)}\to\sH$ maps the solutions of the Schr\"odinger for the Hamiltonian $H(t)$ to those for $\bigh(t)$,
    \be
    U(t,t_0)=\bigrho(t)^{-1}\bigu(t,t_0)\bigrho(t_0).
    \label{HP7}
    \ee
Eqs.~(\ref{HP1}) and (\ref{HP4}) -- (\ref{HP6}) imply
    \be
    O^{\rm(H)}(t)=\bigrho(t_0)^{-1}\bigo^{\rm(H)}(t)\bigrho(t_0).
    \label{Hp7-1}
    \ee
Differentiating both sides of this equation with respect to $t$ and
making use of (\ref{HP3}), (\ref{HP7}), and the identity,
    \[\bigrho(t_0)^{-1}\bigh^{\rm(H)}(t)\bigrho(t_0)-
    iU(t,t_0)^{-1}\bigrho(t)^{-1}\dot\bigrho(t)U(t,t_0)=H^{\rm(H)}(t),\]
which follows from (\ref{xx-28}) and (\ref{HP7}), we arrive at the
Heisenberg equation in the representation $(\sH_{\eta(t)},H(t))$:
    \be
    i\partial_t O^{\rm(H)}(t)=[O^{\rm(H)}(t),H^{\rm(H)}(t)]+i
    U(t,t_0)^{-1}\dot O(t)U(t,t_0).
    \label{HP8}
    \ee

Observe that because $\bigrho(t_0):\sH_{\eta(t_0)}\to\sH$ is a
unitary operator and $\bigo^{\rm(H)}(t):\sH\to\sH$ is Hermitian,
(\ref{Hp7-1}) shows that $O^{\rm(H)}(t)$ acts as a Hermitian
operator in $\sH_{\eta(t_0)}$. This is consistent with the basic
requirement that for an evolving state vector $\psi(t)$,
    \be
    \frac{\br\psi(t),O(t)\psi(t)\kt_{\eta(t)}}{\br\psi(t),\psi(t)\kt_{\eta(t)}}=
    \frac{\br\psi(t_0),O^{(\rm H)}(t)\psi(t_0)\kt_{\eta(t_0)}}{\br\psi(t_0),
    \psi(t_0)\kt_{\eta(t_0)}}.
    \label{HP-z1}
    \ee

Comparing (\ref{HP3}) and (\ref{HP8}), we see that there is no
structural difference between the Heisenberg equations for the
representations $(\sH,\bigh(t))$ and $(\sH_{\eta(t)},H(t))$.

\section{Identification of the energy operator}

The conflict between the unitarity of dynamics and the observability
of the Hamiltonian that appears in the representations of quantum
system with a time-dependent Hilbert space shows that the
Hamiltonian operator appearing in the standard Schr\"odinger
equation does not coincide with the operator associated with the
energy observable in these representations. The distinction between
theses operators seems to disappear when the Hilbert space is
static, simply because we are accustomed to follow the convention of
identifying them. The above conflict provides a clear indication
that this convention is not generally consistent. In the following,
we argue that it is misleading even when the Hilbert space is
time-independent.

Consider a quantum system $\cS$ that is represented using a Hilbert
space $\sH$ with a constant inner product $\br\cdot|\cdot\kt$ and a
Hermitian Hamiltonian operator $\bigh$ acting in $\sH$. The
observables of $\cS$ correspond to the Hermitian operator $\bigo$ acting
in $\sH$. Now, consider a time-dependent unitary operator $\cU(t)$
that maps $\sH$ onto $\sH$. As is well-known, such an operator
induces a quantum analog of a time-dependent classical canonical
transformation. To see this, we recall that $\cU(t)$ induces the
following transformations on the state vectors $\Psi\in\sH$ and the
Hermitian operators $\bigo:\sH\to\sH$:
    \begin{align}
    &\Psi\to\tilde\Psi:=\cU(t)\Psi,
    &&\bigo\to\tilde\bigo:=\cU(t)\,\bigo\,\cU(t)^{-1}.
    \label{trans1}
    \end{align}
These together with the fact that  $\cU(t)^\dagger=\cU(t)^{-1}$
ensure that the expectation values,
$\br\Psi|\bigo\Psi\kt/\br\Psi|\Psi\kt$, are invariant under these
transformations. Therefore we can compute the kinematic properties
of the system at any instant of time using either of $\Psi$ and $\bigo$
or $\tilde\Psi$ and $\tilde\bigo$. The same applies for the dynamical
properties of the system provided that we postulate the following
rule for the transformation of the Hamiltonian
    \be
    \bigh\to\tilde\bigh:=\cU(t)\,\bigh\,\cU(t)^{-1}+i\,\dot{\cU}(t)\,\cU(t)^{-1}.
    \label{trans2}
    \ee
This ensures that $\Psi(t)$ is a solution of the Schr\"odinger
equation of the Hamiltonian $\bigh$ if and only if
$\tilde\Psi(t):=\cU(t)\Psi(t)$ solves  the Schr\"odinger equation
for the Hamiltonian $\tilde\bigh$.

Comparing (\ref{trans1}) and (\ref{trans2}), we see that under
time-dependent quantum canonical transformations, the operators
marking the observables of the system do not transform like the
Hamiltonian operator.\footnote{This is also true about the
transformation property of the observables and the Hamiltonian in
classical mechanics.} In particular, if we employ the convention of
identifying the Hamiltonian $\bigh$ with the energy operator
$\bigh_\ssE$, i.e., set $\bigh_\ssE=\bigh$, we cannot do the same
after we perform the time-dependent quantum canonical transformation
induced by $\cU(t)$; $\bigh\to\tilde\bigh$ while
$\bigh_\ssE\to\tilde\bigh_\ssE=\tilde\bigh-i\,\dot{\cU}(t)\,\cU(t)^{-1}\neq\tilde\bigh$.
This argument shows that we cannot consistently use this convention.
In fact there seems to be no way of determining the energy operator,
if we only know the Hamiltonian operator.

The additional structure that together with the Hamiltonian operator
provide a consistent identification of the energy operator turns out
to have a purely geometric nature \cite{prd-2018}. The subtlety of
dealing with time-dependent Hilbert spaces that we have examined in
the preceding sections provides an important clue for uncovering
this structure. The differential operator $\sD_t$ appearing on the
left-hand side of the modified Schr\"odinger equation
(\ref{mod-sch-eq}) resembles a covariant time derivative with the
term $\bigrho^{-1}\dot\bigrho$ reflecting the contribution of a
local connection (gauge potential). According to (\ref{x4}) and
(\ref{x05}) subtracting this term from the Hamiltonian operator
gives the energy operator. Therefore, it seems that in order to
identify a unique energy operator, we should look for an underlying
vector (or principal) bundle $\cE$ endowed with a connection
\cite{cecile,nakahara,GP-book}. Such a vector bundle has been
constructed in Ref.~\cite{prd-2018} and used to formulate a
geometric extension of quantum mechanics. The standard QM corresponds to
situations where this bundle has a trivial topology. It is however
important to recognize that topologically trivial vector bundles can
possess nontrivial geometries. Indeed, it turns out that the
determination of the energy observable is equivalent to the choice
of a certain geometric structure, namely a metric-compatible
connection, on this vector bundle.

\section{Geometric formulation of quantum dynamics}

\subsection{Vector bundles}

A vector bundle is a manifold $\cE$ equipped with another manifold $M$, a function $\pi$ mapping $\cE$ onto $M$, and a vector space $V$ such that the following conditions hold.
    \begin{itemize}
    \item[-] There are open coordinate patches $\cO_\alpha$ covering $M$ such that the subsets of $\cE$ that are mapped into each of these patches, i.e.,
        \[\cE_\alpha:=\left\{p\in\cE\,|\,\pi(p)\in\cO_\alpha\,\right\},\]
   have the same topological structure as  $O_\alpha\times V$. This means that for each $O_\alpha$, there is a continuous and invertible function $f_\alpha$ with a continuous inverse that maps $\cE_\alpha$ onto $O_\alpha\times V$.
    \item[-] For each $\smR\in M$, the points of $\cE$ that are mapped to $\smR$ by the function $\pi$ form a vector space $V_\ssR$;
    \item[-] For each $\smR\in M$ and $p\in V_\ssR$, let $v$ be the element of $V$ such that $f_\alpha(p)=(\smR,v)$. Then the function,
    \[\phi_{\alpha,\ssR}:V_{\ssR}\to V,\]
that is defined by $\phi_{\alpha,\ssR}(p):=v$ is a vector-space isomorphism, i.e., it is an invertible linear operator mapping $V_\ssR$ onto $V$. In particular, $V_\ssR$ and $V$ are isomorphic vector spaces.
    \end{itemize}
The manifolds $\cE$ and $M$ are called the total and base spaces,
and the vector spaces $V$ and $V_\ssR$ are called the typical fiber
and the fiber over $\smR$, respectively.

The basic motivation for the above definition of a vector bundle is
actually very simple. Consider a pair of coordinate patches,
$\cO_\alpha$ and $\cO_{\tilde\alpha}$, with a nonempty intersection.
Then for each $\smR\in\cO_\alpha\cap\cO_{\tilde\alpha}$, we can use
the so-called transition functions,
	\be
	\biggg_{\tilde\alpha\alpha,\ssR}:=\phi_{\tilde\alpha,\ssR}\circ\phi_{\alpha,\ssR}^{-1},
	\label{gaaR=}
	\ee
to construct a one-to-one correspondence between the points of
$\cO_{\alpha}\times V$ and $\cO_{\tilde\alpha}\times V$:
    \be
    \cO_{\alpha}\times V\ni(\smR,v)
    \xleftrightarrow{~\biggg_{\tilde\alpha\alpha,\ssR}~}
    (\smR,\tilde v)\in \cO_{\tilde\alpha}\times V
    ~~\mbox{if}~~\tilde v=\biggg_{\tilde\alpha\alpha,\ssR}(v).
    \label{glue}
    \ee
This correspondence allows us to reconstruct the total space of the
vector bundle using the knowledge of the patches $\cO_\alpha$ of $M$
and the transition functions $\biggg_{\tilde\alpha\alpha,\ssR}$. To
see this, we associate to each patch $\cO_\alpha$ and
$\smR\in\cO_\alpha$ a vector space $V_{\alpha,\ssR}$ that is an
identical copy of $V$ and suppose that $V_{\alpha,\ssR}$'s with
different $(\alpha,\smR)$ do not intersect, i.e., there is an
isomorphism $\chi_{\alpha,\ssR}:V_{\alpha,\ssR}\to V$, and
$V_{\alpha,\ssR}\cap V_{\alpha',\ssR'}\neq\emp$ if and only if
$\cO_{\alpha}=\cO_{\alpha'}$ and $\smR=\smR'$. We also introduce
    \bea
    \cV_\alpha&:=&\bigcup_{\ssR\in\cO_\alpha}V_{\alpha,\ssR},\nn\\
    \sE_\alpha&:=&\left\{(\smR,v_\alpha)\in \cO_\alpha\times\cV_\alpha~|~v_\alpha\in V_{\alpha,\ssR}~\right\},\nn
    \eea
and note that because $\cE_\alpha$ is an identical copy of
$\cO_\alpha\times V$, we can use $\chi_{\alpha,\ssR}$ to identify
$\cE_\alpha$ with $\sE_\alpha$. This observation together with the
fact that $\cE=\bigcup_\alpha\cE_\alpha$ suggests us to compare
$\cE$ with $\sE:=\bigcup_\alpha\sE_\alpha$. These differ, because if
$\smR\in\cO_\alpha\cap\cO_{\tilde\alpha}$ for some
$\tilde\alpha\neq\alpha$, then to the fiber $V_\ssR$ in $\cE$ there
corresponds two identical copies in $\sE$, namely $V_{\alpha,\ssR}$
and $V_{\tilde\alpha,\ssR}$. This shows that we can obtain $\cE$
from $\sE$ provided that we glue $V_{\alpha,\ssR}$ and
$V_{\tilde\alpha,\ssR}$ along the intersections of the coordinate
patches of $M$. Transition functions $\biggg_{\tilde\alpha
\alpha,\ssR}$ provide the missing gluing rule; we can use them to
introduce the functions,
    \[\check \biggg_{\tilde\alpha \alpha,\ssR}:=\chi_{\tilde\alpha,\ssR}^{-1}\circ \biggg_{\tilde\alpha \alpha,\ssR}\circ \chi_{\alpha,\ssR}:V_{\alpha,\ssR}\to V_{\tilde\alpha,\ssR},\]
and glue $V_{\alpha,\ssR}$ and $V_{\tilde\alpha,\ssR}$ according to
the following prescription:
    \[V_{\alpha,\ssR}\ni (\smR,v_\alpha)~\mbox{is to be glued to}~
          (\smR,v_{\tilde\alpha})\in V_{\tilde\alpha,\ssR}~~{\rm if}~~
          v_{\tilde\alpha}=\check \biggg_{\tilde\alpha \alpha,\ssR}(v).\]

Because the transition functions are automorphisms of $V$, they
belong to a subgroup $G$ of the general linear group $GL(V)$ of all
automorphisms of $V$. The group $G$ is called the structure group of
the vector bundle.

If the fibers of $\cE$ are complex (respectively real) vector
spaces, $\cE$ is called a complex (respectively real) vector bundle.
If, as a manifold, $\cE$ coincides with $M\times V$, it is said to
be a trivial vector bundle. For example $\cE_\alpha$ is a trivial
vector bundle with base space $\cO_\alpha$, because it has the same
topological structure as $\cO_\alpha\times V$. This shows that every
vector bundle is locally trivial, for it can be expressed as the
union of trivial vector bundles.

A smooth function $\psi:M\to\cE$ that maps every point $\smR$ of $M$
to a point in the fiber $V_\ssR$ over $\smR$ is called a global
section of the bundle $\cE$. It turns out that if there are global
sections $\psi_1,\psi_2,\cdots,\psi_\ssN$ such that for each
$\smR\in M$, $\{\psi_1(\smR),\psi_2(\smR),\cdots,\psi_\ssN(\smR)\}$
is a basis of $V_\ssR$, then $\cE$ is a trivial bundle. The converse
is also true if $V$ is an $\smN$-dimensional vector space. For
example, we can always construct such a collection of basis sections
for the vector bundles $\cE_\alpha$. Because the domain of
definition of these sections are not the whole base manifold but
only one of its coordinate patches, namely $\cO_\alpha$, they are
called local sections of $\cE$.

\subsection{Parallel transportation and energy operator}

The geometry of a vector bundle $\sE$ refers to a well-defined
notion of parallel transportation of its points along curves in its
base space $M$. This is achieved by an additional structure called a
``connection.'' We can reduce the problem of defining parallel
transformation along curves in $M$ to that for the segments of the
curve that lie in particular patches of $M$. If we know how do
define the parallel transportation of the points along each of these
segments, we can pass from one patch to the adjacent one using the
transition functions of the bundle. In the following we describe
parallel transportation in a single patch.

Consider a coordinate patch $\cO_\alpha$ of $M$, and identify the
points $\smR$ of $\cO_\alpha$ with its real coordinates
$(\smR^1,\smR^2,\cdots,\smR^d)$. To characterize the points of the
fibers we also introduce a fiber coordinate system. Suppose that $V$
is a finite-dimensional complex vector space. Then without loss of
generality we can identify it with $\C^\ssN$ for some $\smN\in\Z^+$.
Let $B:=\{ e_1, e_2,\cdots, e_\ssN\}$ be the standard basis of
$\C^\ssN$, i.e., $
e_m:=(\delta_{m1},\delta_{m2},\cdots,\delta_{m\ssN})$ where
$\delta_{mn}$ is the Kronecker delta symbol. Because
$\phi_{\alpha,\ssR}:V_\ssR\to V=\C^\ssN$ is an isomorphism,
$\phi^{-1}_{\alpha,\ssR}( e_m)$ form a basis of $V_\ssR$. The
functions $\psi_m:\cO_\alpha\to\cE_\alpha$ defined by
    \be
    \psi_m(\smR):=\phi^{-1}_{\alpha,\ssR}( e_m)
    \label{local-sec}
    \ee
are examples of local sections of $\cE$ that yield a basis of
$V_\ssR$ for $\smR\in\cO_\alpha$, namely 
	\[B_\ssR:=\{\psi_1(\smR),\psi_2(\smR),\cdots,\psi_\ssN(\smR)\}.\]
Given an element $v_\ssR$ of $V_\smR$, we can expand it in this
basis and use the coefficients of this expansion as the coordinates
of $v_\ssR$. In particular, if $\psi:M\to\cE$ is a global section of
$\cE$, there are smooth functions $\Psi_n:\cO_\alpha\to\C$
fulfilling
    \[\psi(\smR)=\sum_{n=1}^\ssN \Psi_n(\smR)\,\psi_n(\smR).\]

We may view the coefficient functions $\Psi_n$ as the components of
a smooth vector-valued function $\Psi:\cO_\alpha\to\C^\ssN$ defined
by
    \[\Psi(\smR):=(\Psi_1(\smR),\Psi_2(\smR),\cdots,\Psi_\ssN(\smR)).\]
Let us now consider a basis transformation,
    \be
    \psi_m(\smR)\to\psi'_m(\smR),
    \label{basis-trans}
    \ee
such that $\psi'_m:\cO_\alpha\to\cE$ are also local sections whose
values form a basis of $V_\ssR$ for each $\smR\in\cO_\alpha$. If
$\Psi'_m:\cO_\alpha\to\C$ are coefficients functions associated with
the expansion of the global section $\psi$ in the basis
$B'_\ssR:=\{\psi'_1(\smR),\psi'_2(\smR),\cdots,\psi'_\ssN(\smR)\}$,
then (\ref{basis-trans}) induces a linear coordinate transformation,
    \be
    \Psi_m(\smR)\to\Psi_m'(\smR)=\sum_{n=1}^\ssN \fg_{mn}(\smR)\Psi_n(\smR),
    \label{coor-trans}
    \ee
where $\fg_{mn}:\cO_\alpha\to\C$ are smooth functions whose values
form the entries of an invertible matrix. Let
$\Psi':\cO_\alpha\to\C^\ssN$ be the analog of $\Psi$ that has
$\Psi'_m$ as its components. Then the coordinate transformation
(\ref{coor-trans}) is equivalent to
    \be
    \Psi(\smR)\to\Psi'(\smR)=\fg(\smR)[\Psi(\smR)],
    \label{gauge-trans-1}
    \ee
where $\fg:\cO_\alpha\to GL(n,\C)$ is a smooth function, and
$GL(\smN,\C):=GL(\C^\ssN)$ is the general linear group of
automorphisms of $\C^\ssN$. The functions $\Psi$ provide local
representations of the global sections $\psi$ in $\cO_\alpha$. In
the applications of vector bundles in particle physics, these
describe the matter fields while the coordinate transformations
(\ref{gauge-trans-1}) correspond to (local) gauge transformations.

Now, consider a smooth curve $\gamma:[t_0,t_1]\to\cO_\alpha$ lying
in $\cO_\alpha$, and identify $\gamma(t)$ with its coordinates
$\smR(t)$. The parallel transportation of a point $\psi_0\in
V_{\ssR(t_0)}$ along $\gamma$ is a  particular assignment of a point
of $V_{\ssR(t)}$ for each $t\in[t_1,t_2]$. This defines a smooth
curve $\Gamma_\ssA:[t_0,t_1]\to\cE_\alpha$. Because
$\pi(\Gamma_\ssA(t))=\gamma(t)$, $\Gamma_\ssA$ is a lift of $\gamma$
from $\cO_\alpha$ to $\cE_\alpha$. It is called the horizontal lift
of $\gamma$. To determine it, we expand $\Gamma_\ssA(t)$ in the basis
$B_{\ssR(t)}$, use $\Psi_n(t)$ to label the coefficients of this
expansion, so that
    \be
    \Gamma_\ssA(t)=\sum_{n=1}^\ssN\Psi_n(t)\psi_n[\smR(t)],
    \nn
    \ee
and identify $\Psi_n(t)$ with the solution of a homogeneous linear
system of first-order differential equations. We can express this
system in the form
    \be
    \sD_t\Psi(t)=0,
    \label{parallel-1}
    \ee
where
    \be
    \sD_t:=\partial_t+i\sum_{a=1}^d \dot\smR^a(t) A_a(\smR(t)),
    \label{connection-1}
    \ee
and $A_a(\smR)$ are linear operators acting in $V=\C^n$, i.e., they
belong to the Lie algebra $\cG\ell(\smN,\C)$ of the group
$GL(\smN,\C)$. In physics literature, they are identified with the
components of a gauge potential.

We can view $A_a(\smR)$ as the value of a smooth function
$A_a:\cO_\alpha\to\cG\ell(n,\R)$ and introduce a
$\cG\ell(n,\C)$-valued one-form $A:=\sum_{a=1}^d A_a d\smR^a$ called
a local connection one-form. Different choices of $A$ determine
different notions of parallel transformation in $\cE_\alpha$.
Demanding that Eq.~(\ref{parallel-1}) preserves its form under a
gauge transformation (\ref{gauge-trans-1}), we are led to the
following gauge transformation rule for local connection one-forms:
$ A\to A'=\fg\,A\,\fg^{-1}-i\fg\, d\fg^{-1}$, where
$d\fg:=\sum_{a=1}^d \partial_{a}\fg\,d\smR^a$ and $\partial_{a}$
stands for partial derivative with respect to $\smR^a$. Let us also
note that the extension of the above procedure for parallel
transformation to curves in $M$ that do not lie in a single local
coordinate patch requires patching together the horizontal lifts
computed in adjacent patches, say $\cO_\alpha$ and
$\cO_{\tilde\alpha}$, at an arbitrary point of the curve that lies
in $\cO_\alpha\cap\cO_{\tilde\alpha}$. We can achieve this provided
that at each $\smR\in\cO_\alpha\cap\cO_{\tilde\alpha}$ the local
connection one-forms $A$ and $\tilde A$, that are respectively
associated with $\cO_\alpha$ and $\cO_{\tilde\alpha}$, are related
via \cite{cecile}
    \be
    \tilde A(\smR)=\biggg_{\alpha\tilde\alpha,\ssR}^{-1} A(\smR)\biggg_{\alpha\tilde\alpha,\ssR}-i\biggg_{\alpha\tilde\alpha,\ssR}^{-1}\,d
    \biggg_{\alpha\tilde\alpha,\ssR}.
    \label{tilde-A=}
    \ee
If we can make a consistent assignment of local connection one-forms
to all the patches $\cO_\alpha$ so that this equation holds in their
intersection, we say that the vector bundle $\cE$ is endowed with a
connection $\cA$.

It is easy to see that we can express Eq.~(\ref{parallel-1}) as the
Schr\"odinger equation,
    \be
    i\partial_t\Psi(t)=H_\ssA(t)\Psi(t),
    \label{sch-eq-A}
    \ee
for a Hamiltonian of the form
    \be
    H_A(t):=\sum_{a=1}^d A_a[\smR(t)]\,\partial_t\smR^a(t),
    \label{HA=}
    \ee
and identify its solution with
    \be
    \Psi(t)=U_\ssA(t,t_0)\Psi(t_0),
    \label{P=UP}
    \ee
where $U_\ssA(t,t_0)$ is the evolution operator for $H_\ssA(t)$.

An important property of the Hamiltonian (\ref{HA=}) is that under
smooth reparametrizations of $t$, i.e., $t\to t'=\tau(t)$ for smooth
monotonically increasing functions $\tau:[t_0,t_1]\to\R$, it
transforms according to $H_\ssA(t)\to
H_\ssA(t')=[\dot\tau(t)]^{-1}H_\ssA(t)$. This implies that such
reparametrizations of time leave the Schr\"odinger equation
(\ref{sch-eq-A}) and hence its solutions invariant. We can express
the time-reparametrization invariance of solutions of
(\ref{parallel-1}) by expressing the time-ordered exponential
yielding $U_\ssA(t,t_0)$ as a path-ordered exponential along
$\gamma$;
    \bea
    U_\ssA(t,t_0)&=&\sT\left\{\exp\int_{t_0}^t\, ds\left[-iH_\ssA(s)\right]\right\}\nn\\
        &=&I+\sum_{\ell=1}^\infty(-i)^\ell
        \int_{t_0}^{t} ds_\ell\int_{t_0}^{s_\ell}ds_{\ell-1}\cdots
        \int_{t_0}^{s_2}ds_1H_\ssA(s_\ell)H_\ssA(s_{\ell-1})\cdots H_\ssA(s_1)\nn\\
        &=&I+\sum_{\ell=1}^\infty(-i)^\ell
        \int_{\ssR(t_0)}^{\ssR(t)} A(\smR_\ell)
        \int_{\ssR(t_0)}^{\ssR_\ell}A(\smR_{\ell-1})
        \cdots \int_{\ssR(t_0)}^{\ssR_2}A(\smR_{1})\nn\\
    &=&
    \sP\left\{\exp\int_{\ssR(t_0)}^{\ssR(t)}\, \left[-i A(\smR)\right]\right\},\nn
    \eea
where $\sT$ and $\sP$ respectively denote time-ordering and
path-ordering operations, and the integrals over the
$\cG\ell(n,\C)$-valued one-forms $A(R_j)$ are to be performed along
the segments of the curve $\gamma$.

The time-reparametrization invariance of the evolution operator for
$H_\ssA$ shows that the dynamics generated by this Hamiltonian  in
the typical fiber $\C^\ssN$ of the bundle $\cE_\alpha$ depends only
on the shape of the curve $\gamma$ and not on how fast this curve is
traversed in time. In other words, it determines a purely
geometrical evolution. Because this evolution yield a
horizontal lift of $\gamma$, we call it a ``horizontal evolution.''

We can also envisage more general lifts of $\gamma$ that are
associated with non-horizontal evolutions in the typical fiber.
These would be determined by Hamiltonians $H(t):\C^\ssN\to\C^\ssN$
whose evolution operator does depend on the parameterization of the
curve $\gamma$. The extreme situation is that of evolutions that
take place in a single fiber of $\cE_\alpha$, i.e., when $\gamma$ is
a constant curve; $\gamma(t)=\smR_0$ for all $t\in[t_0,t_1]$ and
some $R_0\in\cO_\alpha$. In the case, the evolution of a point
$\psi_0\in V_{\ssR_0}$ maps it to
        \be
        \psi_\scE(t):=\sum_{n=1}^\ssN\Psi_n(t)\,\psi_n(\smR_0),
        \label{psi-bundle}
        \ee
where $\Psi_n(t)$ are components of the solution of the
Schr\"odinger equation (\ref{sch-eq}) for a Hamiltonian
$H_\ssE(t):\C^\ssN\to\C^\ssN$. Because in this case $\psi_\scE(t)\in
V_{\ssR_0}$, we call the time-evolution generated by $H_\ssE(t)$ a
``vertical evolution.''

The more general time-reparametrization non-invariant dynamics 
corresponds to an evolution generated by a Hamiltonian of the form,
    \be
    H(t)=H_\ssA(t)+H_\ssE(t).
    \label{H=HH}
    \ee
In this case we can use (\ref{connection-1})and (\ref{HA=}) to
express the Schr\"odinger equation,
    \be
    i\partial_t\Psi(t)=H(t)\Psi(t),
    \label{sch-eq-H}
    \ee
in the form
    \be
    \sD_t\Psi(t)=H_\ssE(t)\Psi(t).
    \label{sch-eq-cov}
    \ee

The modified Schr\"odinger equation (\ref{mod-sch-eq}) that is
proposed in Refs.~\cite{znojil-2008,gong-2013} to circumvent the
conflict between unitarity and the observability of time-dependent
pseudo-Hermitian Hamiltonians is a special case of
(\ref{sch-eq-cov}). If we consider the realistic situations where the
time-dependence of the Hamiltonian and the energy operator is
governed through their dependence on a set of real dynamical control
parameters, which we can identify with coordinates $\smR$ of points
of a parameter space $M$, then $\bigeta=\bigeta(\smR)$,
$\bigrho=\bigrho(\smR)$, and for $\smR=\smR(t)$ we have
$\dot\bigrho=\sum_{a=1}^\ssN\partial_{a}\bigrho \,\dot\smR^a$. With
the help of this relation, we can identify (\ref{covariant}) with the special 
case of (\ref{connection-1}) that is given by the following
choice for the local connection one-form.
    \be
    A=-i\bigrho^{-1}d\bigrho,
    \label{A-special}
    \ee
where $d\bigrho:=\sum_{a=1}^\ssN \partial_{a}\bigrho\, d\smR^a$. It
is this choice that identifies the energy observable $H_\ssE$ with
the ``Hamiltonian'' for the modified Schr\"odinger equation
(\ref{mod-sch-eq}).

The above analysis suggests that we can keep using the term
``Hamiltonian'' for the generator of time evolutions $H$ in the
Schr\"odinger equation (\ref{sch-eq-H}), and identify the energy
operator with the generator of vertical evolutions $H_\ssE$. It is
then clear that the knowledge of $H$ is not sufficient to determine
$H_\ssE$ unless we also know $H_\ssA$. Given that the latter is
uniquely determined by the connection one-form $A$, we are led to a
geometric formulation of quantum dynamics where we can identify
the evolution of state vectors with certain trajectories in a trivial
vector bundle $\cE_\alpha$ endowed with a local connection-one form
$A$. Each such trajectory is a lift of a curve of control parameters
of the system. It is determined by the choice of $A$ and the 
energy operator $H_\ssE$. We can relate the
latter with an assignment of a linear operator $\fH(\smR):V_\ssR\to
V_\ssR$ to each $\smR\in\cO_\alpha$, because we can specify $H_\ssE$
in the form
    \be
    H_\ssE(t)=\phi_{\alpha,\ssR(t)}\circ\fH(\smR(t))\circ\phi_{\alpha,\ssR(t)}^{-1}.
    \label{EP-local}
    \ee
We can view $\fH$ as a function mapping $\cO_\alpha$ into another
vector bundle which we describe after we elucidate the notion of
``observable'' in our vector bundle setting for QM.

We end this subsection by stressing that the choice (\ref{A-special})
for $A$ is not dictated by any basic physical principle. This choice
follows from the requirement of identifying the Hamiltonian
$\bigh(t)$ with the energy operator. But as we discussed above, this
requirement violates the invariance of expectation values of the
energy observable under time-dependent quantum canonical
transformations.

\subsection{Hermitian vector bundles, unitarity, and observables}

If each of the fibers $V_\ssR$ of a complex vector bundle $\sE$ is
equipped with an inner product $\br\cdot,\cdot\kt_\ssR$,
we call $\cE$ a Hermitian vector bundle.
This inner product makes the fibers of $\sE$ into an inner-product
space. For cases where the fibers are finite-dimensional, they are
Hilbert spaces parameterized by the points $\smR$ of $M$.\footnote{If
the fibers $V_\ssR$ are infinite-dimensional separable Hilbert spaces, $\sE$ is 
called a Hilbert bundle. These turn out to be topologically trivial
\cite{kuiper,sen}, but they may possess nontrivial geometries.}

If a Hermitian vector bundle is endowed with a connection, parallel
transportations of a pair of points belonging to a fiber may change
their inner product. There are however a special class of connections 
on Hermitian bundles where this does not happen,
i.e., parallel transportation along all curves preserves the inner
product. Such a connection is called a metric-compatible or simply a
metric connection.

Let us fix a coordinate patch $\cO_\alpha$ of $M$ and use the local
sections $\psi_m:\cO_\alpha\to\cE_\alpha$ defined by
(\ref{local-sec}) together with the inner product on the fibers of
$\cE_\alpha$ to construct an inner product on $V=\C^\ssN$ as
follows.

\begin{figure}[ht]
    \begin{center}
    \includegraphics[scale=.25]{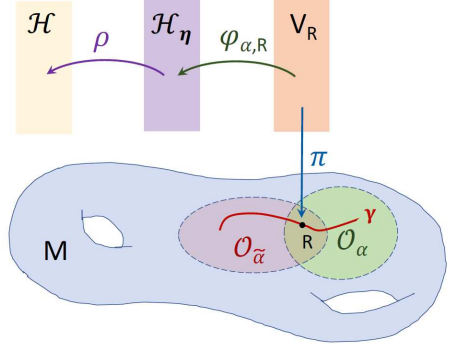}
    \caption{Schematic diagram representing the base space $M$ of the vector bundle $\cE$, a curve $\gamma$ in $M$, a pair of intersecting coordinate patches $\cO_\alpha$ and $\cO_{\tilde\alpha}$ of $M$ that cover $\gamma$. $R$ is a point in $\cO_\alpha\cap\cO_{\tilde\alpha}$. The function $\pi:\cE\to M$ is the bundle projection map that maps the fiber $V_\ssR$ over $R$ to $R$, i.e., $V_\ssR=\pi^{-1}(\{R\})$. $\sH_\eta$ and $\sH$ are respectively the typical fiber $\C^\ssN$ endowed with the inner products $\br\cdot,\cdot\kt_\eta$ and the Euclidean inner product $\br\cdot|\cdot\kt$. The isomorphisms $\varphi_{\alpha,R}:V_\ssR\to\sH_\eta$ and $\bigrho:\sH_\eta\to\sH$ are unitary operators.}
    \label{fig1}
    \end{center}
    \end{figure}

First, we introduce
    \be
    \eta_{mn}(\smR):=\br\psi_m(\smR),\psi_n(\smR)\kt_\ssR,
    \label{eta-def1}
    \ee
and identify $\bigeta(\smR):\C^\ssN\to \C^\ssN$ and $\br\cdot,\cdot\kt_{\eta(\ssR)}:\C^\ssN\times\C^\ssN\to\C$ with the linear operator and inner product
defined by
    \begin{align}
    &\bigeta(\smR)w:=\sum_{m=1}^\ssN \eta_{mn}(\smR)w_n,
    &&\br\cdot,\cdot\kt_{\eta(\ssR)}:=\br\cdot|\bigeta(\smR)\cdot\kt,
    \label{eta-def2}
    \end{align}
where $w:=(w_1,w_2,\cdots,w_\ssN)$ is an arbitrary element of
$\C^\ssN$, and $\br\cdot|\cdot\kt$ is the Euclidean inner product on
$\C^\ssN$. Then, for every $v:=(v_1,v_2,\cdots,v_\ssN)\in\C^\ssN$,
we have
    \bea
    \br v,w\kt_{\eta(\ssR)}&=&\br v|\bigeta(\smR)w\kt=
    \sum_{m,n=1}^\ssN v_m^*\eta_{mn}(\smR)w_n\nn\\
    &=&\sum_{m,n=1}^\ssN v_m^*w_n\br\psi_m(\smR),\psi_n(\smR)\kt_\ssR \nn\\
    &=&\sum_{m,n=1}^\ssN v_m^*w_n
    \br\phi_{\alpha,\ssR}^{-1}( e_m),\phi_{\alpha,\ssR}^{-1}( e_n)\kt_\ssR\nn\\
    &=&\br\phi_{\alpha,\ssR}^{-1}(v),\phi_{\alpha,\ssR}^{-1}(w)\kt_\ssR.
    \label{x51}
    \eea
This calculation shows that $\br\cdot,\cdot\kt_{\eta(\ssR)}$ is a
genuine inner product on $\C^\ssN$, and $\bigeta$ is a metric
operator acting in the Hilbert space
$\sH:=(\C^\ssN,\br\cdot|\cdot\kt)$. Furthermore, (\ref{x51}) implies
that if we use $\sH_{\eta(\ssR)}$ to denote the Hilbert space
$(\C^\ssN,\br\cdot,\cdot\kt_{\eta(\ssR)})$, the isomorphisms
$\phi_{\alpha,\ssR}:V_\ssR\to \sH_{\eta(\ssR)}$ are unitary
operators. See Fig~\ref{fig1} for a schematic representation of the 
related mathematical constructs.

Next, suppose that $\cE$ is provided with a connection $\cA$, and
$A$ is the corresponding local connection one-form on $\cE_\alpha$.
Let $\gamma:[t_0,t_1]\to\cO_\alpha$ be a smooth curve, $\smR(t)$
label the coordinates of  $\gamma(t)$, and $\phi(t)$ and $\psi(t)$
be elements of $V_{\ssR(t)}$ that are respectively obtained by the
parallel transportation of points $\phi_0$ and $\psi_0$ of
$V_{\ssR(t_0)}$ along $\gamma$. By definition, $\cA$ is a metric
connection if for all choices of $\cO_\alpha$, $\gamma$, $\phi_0$,
and $\psi_0$,
    \be
    \br \phi(t),\psi(t)\kt_{\ssR(t)}=\br \phi_0,\psi_0\kt_{\ssR(t_0)}.
    \label{metric-conn-1}
    \ee
If $\Phi_n(t)$ and $\Psi_n(t)$ are the coefficients of the expansion
of $\phi(t)$ and $\psi(t)$ in the local sections $\psi_n(\smR(t))$,
so that
    \begin{align*}
    &\phi(t)=\sum_{n=1}^\ssN\Phi_n(t)\psi_n(\smR(t)),
    &&\psi(t)=\sum_{n=1}^\ssN\Psi_n(t)\psi_n(\smR(t)),
    \end{align*}
and $\Phi:=(\Phi_1,\Phi_2,\cdots,\Phi_\ssN)$ and $\Psi:=(\Psi_1,\Psi_2,\cdots,\Psi_\ssN)$, we can use (\ref{eta-def1}) and (\ref{eta-def2}) to express (\ref{metric-conn-1}) as
    \be
    \br\Phi(t),\Psi(t)\kt_{\eta(t)}=\br\Phi(t_0),\Psi(t_0)\kt_{\eta(t_0)},
    \label{xx-247}
    \ee
where $\bigeta(t):=\bigeta(\smR(t))$. Eqs.~(\ref{P=UP}) and
(\ref{xx-247}) show that the evolution operator $U_\ssA(t,t_0)$
associated with the Hamiltonian $H_\ssA(t)$ acts in the Hilbert
space $\sH_{\eta(t)}$ as a unitary operator. That is horizontal
evolutions defined by a metric connection in $H_\ssA(t)$ are
unitary. In particular, $H_\ssA(t)$ satisfies (\ref{xx0}).
Equivalently,
    \be
    H_\ssA(t)^\dagger=\bigeta(t)H_\ssA(t)\bigeta(t)^{-1}+i\dot\bigeta(t)\bigeta(t)^{-1}.
    \label{HA-unitary}
    \ee

Now, consider a general lift of the curve $\gamma$ that is
determined by (\ref{H=HH}), (\ref{sch-eq-H}), and
(\ref{sch-eq-cov}). Then the evolution operator $U(t,t_0)$ defines a
unitary operator acting in $\sH_{\eta(t)}$ if and only if the
Hamiltonian $H(t)$ satisfies (\ref{xx0}). In view of
(\ref{HA-unitary}), we can express this condition in the form
    \be
    H_\ssE(t)^\dagger=\bigeta(t)H_\ssE(t)\bigeta(t)^{-1},
    \label{HE-unitary}
    \ee
i.e., $H_\ssE(t)$ acts as an $\bigeta$-pseudo-Hermitian operator in
$\sH$ and as a Hermitian operator in $\sH_{\eta(t)}$. As a result,
its expectation values are real provided that we compute them using
the inner product (\ref{eta-def2}). This suggests that we can safely
identify it with an observable of a unitary quantum system $\cS$
that is represented by the pair $(\sH_{\eta(t)},H(t))$ and call it
the energy operator.

We can represent the quantum system $\cS$ also using
$\big(\sH,\bigh(t)\big)$, where $\bigh(t)$ is given by
(\ref{xx-28}). In view of this relation and (\ref{H=HH}), $\bigh(t)$
admits the decomposition:
    \[\bigh(t)=\bigh_\ssA(t)+\bigh_\ssE(t),\]
where
    \begin{align}
    &\bigh_\ssA(t):=\bigrho(t)H_\ssA(t)\bigrho(t)^{-1}+i\dot\bigrho(t)\bigrho(t)^{-1},
    \label{hA=}\\
    &\bigh_\ssE(t):=\bigrho(t)H_\ssE(t)\bigrho(t)^{-1},
    \label{hE=}
    \end{align}
and $\bigrho(t):=\sqrt{\bigeta(t)}$. It is not difficult to show
that both $\bigh_\ssA(t)$ and $\bigh_\ssE(t)$ act as Hermitian
operators in $\sH$. According to (\ref{hE=}), $\bigh_\ssE(t)$ is the
energy operator in this representation. Let us also note that the
special choice (\ref{A-special}) for the local connection one form
$A$ implies $H_\ssA(t)=-i\bigrho(t)^{-1}\dot\bigrho(t)$.
Substituting this equation in (\ref{hA=}), we find
$\bigh_\ssA(t)=0$. Therefore, it is only for this choice of $A$ that
$\bigh(t)$ coincides with the energy operator $\bigh_\ssE(t)$.

Next, we recall that the operator
$\phi_{\alpha,\ssR}:V_\ssR\to\sH_{\eta(\ssR)}$ is unitary.
Therefore, we can use it to construct another representation of the
quantum system $\cS$ where the state vectors at time $t$ belong to
the fiber $V_{\ssR(t)}$, the observables measured at this time are
given by Hermitian operators $\fO:V_{\ssR(t)}\to V_{\ssR(t)}$, and
the dynamics corresponds to the lifts of the curve $\gamma$ traced
by the control parameters $\smR$. In particular, the evolving states
$\psi(t)$ are given by (\ref{psi-bundle}) with $\Psi_n$ being
components of a solution $\Psi$ of (\ref{sch-eq-cov}). It is not
difficult to see that
    \be
    \psi(t)=\phi_{\alpha,\ssR(t)}^{-1}(\Psi(t)).
    \label{psi-phi-psi}
    \ee
Solving this equation for $\Psi(t)$ and substituting the result in
(\ref{sch-eq-cov}), we can identify $\psi:[t_0,t]\to\cE_\alpha$ with
a solution of the evolution equation,
    \be
    iD_t\psi(t)=\fH(t)\psi(t),
    \label{cov-evo-eq}
    \ee
where
    \begin{align}
    &D_t:= \phi_{\alpha,\ssR(t)}^{-1}\circ\sD_t\circ \phi_{\alpha,\ssR(t)}
    \end{align}
is called the covariant time-derivative corresponding to the metric
connection on $\cE$, and
    \begin{align}
    &\fH(t):=\phi_{\alpha,\ssR(t)}^{-1}\circ H_\ssE(t)\circ \phi_{\alpha,\ssR(t)}
    \end{align}
is a Hermitian operator acting in $V_{\ssR(t)}$ that represents the
energy observable of $\cS$.

The existence of a representation of $\cS$ that uses the fibers of
$\cE_\alpha$ as the Hilbert space of state vectors and identifies
the Hermitian operators acting in these fibers with the observables
suggests a natural extension where the possibly nontrivial Hermitian
vector bundle $\cE$ plays the role of its trivial subbundle
$\cE_\alpha$. This leads to a proposal for a geometric extension of
quantum mechanics that we examine in the next section.

\section{Geometric extension of quantum mechanics}

Any attempt at extending QM must address both its kinematic and
dynamical aspects.\footnote{By kinematic aspects, we mean the 
definition of states, observables, and the meaning and implications
of observing an observable when the system is in a given state. By 
dynamical aspects, we mean the prescription according to which the 
time-evolution of the states or observables of the system are 
determined.} In  particular, it should clarify how such an
attempt affects or alters the projection axiom. Obviously, the most
conservative approach is to make sure this axiom holds in a more
general setting. In trying to extend the description of a quantum
system using a trivial Hermitian vector bundle to situations that
the bundle has a nontrivial topology, this can be easily achieved,
for a measurement of an observable takes place at a single instant
of time. This observation together with the developments we have
reported above lead to a natural geometric extension of quantum
mechanics (GEQM) that we describe in the sequel.

The postulates of GEQM involve another vector bundle which we label
by $\bigfu(\cE)$. This is a real vector bundle with base space $M$.
Its fiber $\fu_\ssR$ over the point $\smR\in M$ is the real vector
space of Hermitian operators acting in the fiber $V_\ssR$ of $\cE$.
Its typical fiber is the vector space of Hermitian operators acting
in $\C^\ssN$, which we can identify with the Lie algebra $\fu(\smN)$
of the unitary group $U(\smN)$.\footnote{Note that we can express the 
elements of $U(\smN)$ in the form $e^{i\fX}$ where $\fX$ is an 
$\smN\times\smN$ Hermitian matrix. Therefore, the elements of the Lie algebra 
$\fu(\smN)$ are of the form $i\fX$. $\fu(\smN)$ has the structure of a real vector 
space, because it is closed under matrix addition and scalar multiplication of 
matrices by real numbers. In physics literature, $\fu(\smN)$ is identified with the 
real vector space of $\smN\times\smN$ Hermitian matrices, because as real 
vector spaces they are isomorphic.} The transition functions
$\bigfg_{\alpha\tilde\alpha,\ssR}:\fu(\smN)\to\fu(\smN)$ of
$\bigfu(\cE)$ are given by the following
relations \cite{prd-2018}.
        \begin{align}
        &\bigfg_{\tilde\alpha\alpha,\ssR}(\bigo):=
        \cG_{\tilde\alpha\alpha,\ssR}\bigo\,
        \cG_{\tilde\alpha\alpha,\ssR}^{-1},&&
        \cG_{\tilde\alpha\alpha,\ssR}:=\tilde\bigrho(\smR)\,
        \biggg_{\tilde\alpha\alpha,\ssR}\,\bigrho(\smR)^{-1},
        \label{transition-u}
        \end{align}
where $\alpha$ and $\tilde\alpha$ label pairs of intersecting
coordinates charts, $\smR$ belongs to their intersection,
$\bigrho(\smR)=\sqrt{\bigeta(\smR)}$,
$\tilde\bigrho(\smR)=\sqrt{\tilde\bigeta(\smR)}$,
$\bigeta(\smR):\C^\ssN\to\C^\ssN$ is the metric operator associated
with the coordinate chart $\cO_\alpha$, which we introduced in
Subsec.~6.3, $\tilde\bigeta(\smR)$ is its analog for the coordinate
chart $\cO_{\tilde\alpha}$, and $\biggg_{\alpha\tilde\alpha,\ssR}$
are the transition functions of $\cE$.

Having introduced $\bigfu(\cE)$, we can present the postulates of
GEQM as follows.
    \begin{itemize}

    \item A quantum system $\cS$ is determined by a complex Hermitian vector bundle $\cE$ endowed with a metric connection $\cA$, a global section $\fH:M\to\bigfu(\cE)$ of the vector bundle $\bigfu(\cE)$, and a smooth parameterized curve $\gamma:[t_0,t_1]\to M$, where the parameter of $\gamma$ is time, $[t_0,t_1]$ is the time interval in which we wish to describe the system, and $M$ is the base space of $\cE$ whose points correspond to a collection of classical external control parameters.

    \item The (pure) states of $\cS$ at a time $t$ are given by one-dimensional subspaces (rays) of the fiber $V_{\ssR(t)}$ of $\cE$, where $\smR(t)$ labels the value of $\gamma$ at $t$. These are uniquely determined by nonzero elements of $V_{\ssR(t)}$ which we identify with the state vectors of $\cS$ at time $t$.

    \item The observables of $\cS$ are represented by global sections $\fO:M\to\bigfu(\cH)$ of $\bigfu(\cE)$. For a measurement of $\fO$ at time $t$, one implements von-Neumann's projection axiom for the operator $\fO(\smR(t))$, which acts as a Hermitian operator in $V_{\ssR(t)}$. In particular, if the system is in the state given by a state vector $\psi\in V_{\ssR(t)}$, the measurement yields a reading that is an eigenvalue $\omega(t)$ of $\fO(\smR(t))$ and causes an abrupt change of the state of the system to one given by an eigenvector of $\fO(\smR(t))$ with eigenvalue $\omega(t)$.   The probability of reading $\omega(t)$ and the expectation value of $\fO(\smR(t))$ are computed using the textbook prescription with $V_{\ssR(t)}$ and $\fO(\smR(t))$ respectively playing the roles of the Hilbert space and the operator representing the observable.

    \item The evolution of the state vectors $\psi(t)$ are determined by the covariant Schr\"odinger equation,
        \be
        iD_t\psi(t)=\fH(\smR(t))\psi(t),
        \label{cov-evo-eq2}
        \ee
where $D_t$ is the covariant time-derivative defined by the
connection, and $\fH$ is the global section of $\bigfu(\cE)$ that
represents the energy observable.

    \end{itemize}
It is not difficult to check that whenever the curve $\gamma$ lies
in a single coordinate patch of $\cE$, we can describe the system
using $\cE_\alpha$. In this case we recover the representation of
the system we outlined in Subsec.~6.3. In particular, we can represent
the system in terms of the Hilbert space $\sH$ and the Hamiltonian
$\bigh(t)$ using the standard rules of QM. This shows that GEQM
reduces to QM locally. The same is the case if $\cE$ happens to be a
trivial bundle. In general, however, $\cE$ is nontrivial, and we
find an extension of QM. At present the physical implications
of the structural differences between GEQM and QM are not known.

It is a well-known mathematical fact that whenever the typical fiber
of a vector bundle $\cE$ is an infinite-dimensional Hilbert space,
it is necessarily trivial \cite{kuiper,sen}.  This suggests that
GEQM and QM are different only for systems with finite-dimensional
state spaces.\footnote{For a specific example of a class of toy
models with two-dimensional state spaces see \cite{prd-2018}.}

The assertion that GEQM and QM coincide for situations where $\cE$
is trivial may seem as a negative result, but we should realize that
topologically trivial vector bundles can possess nontrivial
geometries. This reveals a hidden geometric aspect of QM that is
directly linked with the problem of identifying the energy operator.

\section{Heisenberg picture of dynamics in GEQM}

In the preceding section we have offered a description of GEQM in which the state vectors undergo dynamical evolutions. For a
given observable represented by a global section
$\fO:M\to\bigfu(\cH)$ of $\bigfu(\cH)$, the expectation value of $\fO$ for a measurement conducted at time $t$ is given by
    \[\frac{\br\psi(t),\fO(\smR(t))\psi(t)\kt_{\ssR(t)}}{
    \br\psi(t),\psi(t)\kt_{\ssR(t)}},\]
where $\br\cdot,\cdot\kt_{\ssR(t)}$ is the inner product of the
fiber $V_{\ssR(t)}$ and $\psi(t)\in V_{\ssR(t)}$ is the state vector
at time $t$.

Suppose that the curve $\gamma$ lies in a single coordinate patch $\cO_\alpha$ of $M$. Then we can use the unitary
transformation $\phi_{\alpha,R}:V_\alpha\to\sH_{\eta(\ssR)}$ to
introduce the operator,
    \be
    O(\smR):=\phi_{\alpha,R}\,\fO(\smR)\phi_{\alpha,R}^{-1},
    \label{HP-O}
    \ee
which acts as a Hermitian operator in $\sH_{\eta(\ssR)}$. Let us also
recall that we determine $\psi(t)$ from
$\Psi(t):=\phi_{\alpha,R(t)}\psi(t)$ and that $\Psi(t)$
satisfies the Schr\"odinger equation defined by the Hamiltonian
operator $H(t)$ in the Hilbert space $\sH_{\eta(t)}$, where
$\bigeta(t):=\bigeta(\smR(t))$.

Because $\phi_{\alpha,R}:V_\alpha\to\sH_{\eta(\ssR)}$ is unitary,     \bea
    \frac{\br\psi(t),\fO(\smR(t))\psi(t)\kt_{\ssR(t)}}{
    \br\psi(t),\psi(t)\kt_{\ssR(t)}}&=&
    \frac{\br\phi_{\alpha,R(t)}\psi(t),
    \phi_{\alpha,R(t)}\,\fO(\smR(t))\psi(t)\kt_{\eta(\ssR(t))}}{
    \br\phi_{\alpha,R(t)}\psi(t),\phi_{\alpha,R(t)}\psi(t)\kt_{\eta(\ssR(t))}}
    \nn\\
    &=&\frac{\br\Psi(t),
    O(t)\Psi(t)\kt_{\eta(t)}}{
    \br\Psi(t),\Psi(t)\kt_{\eta(t)}}\nn\\
    &=&\frac{\br\Psi(t_0),
    O^{\rm(H)}(t)\Psi(t_0)\kt_{\eta(t_0)}}{
    \br\Psi(t_0),\Psi(t_0)\kt_{\eta(t_0)}}\nn\\
    &=&\frac{\br\phi_{\alpha,R(t_0)}\psi(t_0),
    O^{\rm(H)}(t)\phi_{\alpha,R(t_0)}\psi(t_0)
    \kt_{\eta(\ssR(t_0))}}{
    \br\phi_{\alpha,R(t_0)}\psi(t_0),\phi_{\alpha,R(t_0)}\psi(t_0)
    \kt_{\eta(\ssR(t_0))}}\nn\\
    &=&\frac{\br\psi(t_0),\fO^{\rm(H)}(t)\psi(t_0)\kt_{\ssR(t_0)}}{
    \br\psi(t_0),\psi(t_0)\kt_{\ssR(t_0)}},
    \label{s8-1}
    \eea
where we have used (\ref{HP-z1}) and (\ref{HP-O}), set $O(t):=O(\smR(t))$, and introduced: 
    \be
    \fO^{\rm(H)}(t):=\phi_{\alpha,R(t_0)}^{-1}O^{\rm(H)}(t)\,
    \phi_{\alpha,R(t_0)}.
    \label{s8-2}
    \ee
This is a Hermitian operator acting in $V_{\ssR(t_0)}$, i.e., it belongs to $\fu_{\ssR(t_0)}$. In view of (\ref{HP5}), we can express it in the form,
    \be
    \fO^{\rm(H)}(t)=\fU(t,t_0)^{-1}\fO(\smR(t))\,\fU(t,t_0),
    \label{s8-3}
    \ee
where $\fU(t,t_0):V_{\ssR(t_0)}\to V_{\ssR(t)}$ is the linear
operator defined by
    \be
    \fU(t,t_0):=\phi_{\alpha,\ssR(t)}^{-1}U(t,t_0)\phi_{\alpha,\ssR(t_0)},
    \label{s8-fU-def}
    \ee
and $U(t,t_0)$ is the evolution operator for the Hamiltonian $H(t)$.

It is easy to see that $\psi(t)=\fU(t,t_0)\psi(t_0)$. This together
with (\ref{s8-1}) and (\ref{s8-3}) suggest identifying
$\fO^{\rm(H)}(t)$ with the Heisenberg-picture operator associated
with the observable represented by the global section $\fO$.
According to (\ref{s8-2}), we can identify $\fO^{\rm(H)}(t)$ with
the solution of the Heisenberg equation (\ref{HP8}) that satisfies 
the initial condition, 
	\be
	O^{\rm(H)}(t_0):=O(t_0)=\phi_{\alpha,R(t_0)}\fO(\smR(t_0))\phi_{\alpha,R(t_0)}^{-1}.
	\label{s8-ini-condi}
	\ee

If the curve $\gamma:[t_0,t_1]\to M$ of the parameters of the system does not lie in a single coordinate patch, we can dissect it into segments belonging to coordinate patches. We can then integrate
(\ref{HP8}) to determine $O^{\rm(H)}(t)$ and $\fO^{\rm(H)}(t)$ for
each segment and connect the solutions using the appropriate
transition functions. To see the details of this procedure, suppose that $\gamma$ consists of segments $\gamma_0:[t_0,\tilde t_0]\to\cO_\alpha$ and $\gamma_1:[\tilde t_0,t_1]\to\cO_{\tilde\alpha}$ where $\cO_\alpha$ and $\cO_{\tilde\alpha}$ are coordinate patches of $M$ with $\smR(\tilde t_0)\in \cO_{\alpha}\cap \cO_{\tilde\alpha}$, i.e.,
	\[\gamma(t)=\left\{\begin{array}{ccc}
	\gamma_0(t) & {\rm for} & t\in[t_0,\tilde t_0],\\
	\gamma_1(t) & {\rm for} & t\in(\tilde t_0,t_1],\end{array}\right.\]
and $\gamma_0(\tilde t_0)=\gamma_1(\tilde t_0)$. Then an initial state vector $\psi(t_0)\in V_{\ssR(t_0)}$ evolves according to
	\be
	\psi(t)=\left\{\begin{array}{ccc}
	\fU(t,t_0)\psi(t_0) & {\rm for} & t\in[t_0,\tilde t_0],\\
	\tilde\fU(t,\tilde t_0)\fU(\tilde t_0,t_0)\psi(t_0)
	& {\rm for} & t\in(\tilde t_0,t_1],\end{array}\right.
	\label{s8-psi=}
	\ee
where $\fU(t,t_0)$ and $\tilde\fU(t,\tilde t_0)$ are respectively given by 
(\ref{s8-fU-def}) and 
    \be
    \tilde\fU(t,\tilde t_0):=\phi_{\tilde\alpha,\ssR(t)}^{-1}\tilde U(t,\tilde t_0)\phi_{\tilde \alpha,\ssR(\tilde t_0)},
    \nn
    \ee
$\tilde U(t,\tilde t_0)$ is the evolution operator associated with the Hamiltonian, 
	\[\tilde H(t):=\tilde H_{\tilde A}(t)+\tilde H_{\sE}(t)=
	\sum_{a=1}^d\tilde A_a[\smR(t)]\dot\smR^a(t)+
	\phi_{\tilde\alpha,\ssR(t)}\fH(\smR(t))\phi_{\tilde
	\alpha,\ssR(t)}^{-1},\]
and the initial time $\tilde t_0$, and $\tilde A_a$ are component of the local connection one-form $\tilde A$ in the patch $\cO_{\tilde\alpha}$ which fulfills (\ref{tilde-A=}).

Eq.~(\ref{s8-psi=}) suggests that the Heisenberg-picture operator $\fO^{\rm(H)}(t):V_{\ssR(t_0)}\to V_{\ssR(t_0)}$ is to be given by (\ref{s8-3}) for $t\in[t_0,\tilde t_0]$, and by
	\bea
	\fO^{\rm(H)}(t)&:=&
	[\tilde\fU(t,\tilde t_0)\fU(\tilde t_0,t_0)]^{-1}
	\fO(\smR(t))\,\tilde\fU(t,\tilde t_0)\fU(\tilde t_0,t_0),
	\label{s8-11}
	\eea
for $t\in(\tilde t_0,t_1]$. Note also that	
	\be
	\tilde\fU(t,\tilde t_0)\fU(\tilde t_0,t_0)=
	\phi_{\tilde\alpha,\ssR(t)}^{-1}\tilde U(t,\tilde t_0)
	\,\biggg_{\tilde\alpha\alpha,\ssR(\tilde t_0)}U(\tilde t_0,t_0)
	\phi_{\alpha,\ssR(t_0)}.
	\label{s8-12}
	\ee
	
It is clear that for $t\in[t_0,\tilde t_0]$, the operator $O^{\rm (H)}(t)$ given by (\ref{HP5}) satisfies (\ref{HP8}). To determine the analog of (\ref{HP8}) for $t\in[\tilde t_0,t_1]$, we let
	\be
	O^{\rm(H)}(t):=\phi_{\alpha,\ssR(t_0)}\fO^{\rm(H)}(t)\phi_{\alpha,\ssR(t_0)}^{-1},\nn
	\ee
and use (\ref{s8-11}) and (\ref{s8-12}) to show that, for $t\in[\tilde t_0,t_1]$,
	\be
	O^{\rm(H)}(t)=U(\tilde t_0,t_0)^{-1}\biggg_{\tilde\alpha \alpha,\ssR(t_0)}^{-1}\tilde O^{\rm(H)}(t)\biggg_{\tilde\alpha \alpha,\ssR(t_0)}
	U(\tilde t_0,t_0),
	\label{s8-13}
	\ee
where $\tilde O^{\rm(H)}(t):=\tilde U(t,\tilde t_0)^{-1}\tilde O(t)
	\tilde U(t,\tilde t_0)$, $\tilde O(t):=\tilde O(\smR(t))$, and
	\begin{align}
	&\tilde O(\smR):=\phi_{\tilde\alpha,\ssR}\,\fO(\smR)
	\phi_{\tilde\alpha,\ssR}^{-1}.
	\label{s8-21}
	\end{align} 		
Pursuing a similar approach as the one leading to (\ref{HP8}), we can show that $\tilde O^{\rm(H)}(t)$ satisfies the Heisenberg equation,
	\be
    	i\partial_t \tilde O^{\rm(H)}(t)=[\tilde O^{\rm(H)}(t),
	\tilde H^{\rm(H)}(t)]+i
    	\tilde U(t,\tilde t_0)^{-1}\dot{\tilde O}(t)\tilde U(t,\tilde t_0),
    	\label{s8-14}
    	\ee
and the initial condition $\tilde O^{\rm(H)}(\tilde t_0):=\tilde O(\tilde t_0)$. According to (\ref{s8-13}), this implies that $O^{\rm(H)}(t)$  satisfies (\ref{s8-14}) and the initial condition: 
	\bea
	O^{\rm(H)}(\tilde t_0)&=&
	U(\tilde t_0,t_0)^{-1}\biggg_{\tilde\alpha \alpha,\ssR(t_0)}^{-1}\tilde O(\tilde t_0)\biggg_{\tilde\alpha \alpha,\ssR(t_0)}
	U(\tilde t_0,t_0)\nn\\
	&=& U(\tilde t_0,t_0)^{-1} O(\tilde t_0)U(\tilde t_0,t_0),
	\label{s8-27}
	\eea
where we have employed (\ref{gaaR=}), (\ref{HP-O}), and (\ref{s8-21}). Notice that (\ref{s8-27}) is consistent with the fact that for $t\in[t_0,\tilde t_0]$, $O^{\rm(H)}(t)$ satisfies (\ref{HP5}). This in turn shows that $O^{\rm(H)}(t)$ traces a smooth curve in the Hilbert space $\sH_{\eta(\ssR(t_0))}$. 

The procedure we have outlined for the cases where $\gamma$ consists of a pair of segments each contained in a  coordinate patch trivially extends to situations where it consists of an arbitrary number of such segments.

\section{Concluding remarks} 

Pseudo-Hermitian operators were initially considered in an attempt to provide a mathematically more careful assessment of some of the claims made by proponents of the importance of $\cP\cT$-symmetry, \cite{p1,p2}. This clarified a number of issues of basic importance such as the spectral consequences of antilinear symmetries \cite{p3}, the idea of reviving the Hermiticity of certain non-Hermitian operators by modifying the inner product of the Hilbert space \cite{p1,jmp-2003,jpa-2003}, and a consistent definition of observables for $\cP\cT$-symmetric systems \cite{jpa-2004,cjp-2004}. These developments involved considering time-independent pseudo-Hermitian Hamiltonian operators and led to various applications of these operators \cite{review}.

The study of time-dependent pseudo-Hermitian Hamiltonian operators was initially motivated by certain basic problems of quantum cosmology \cite{cqg-2003,ap-2004}. An important outcome of this study is a curious conflict between the unitarity of dynamics generated by such Hamiltonians and their observability \cite{plb-2007}. This conflict has a more general domain of validity, for it applies to every quantum system whose state space is time-dependent. A proper resolution of this conflict calls for a more careful examination of the notion of energy operator for such systems. In this article, we have provided a geometric setting for addressing this issue, described the geometric meaning of the energy operator as the generator of vertical evolutions in a Hermitian vector bundle. A by-product of this approach is a consistent geometric extension of quantum mechanics. We have offered a general description of this extension and outlined the Heisenberg-picture formulation of its dynamical aspects.\vspace{12pt}

\noindent{\bf Funding:} This work has been supported by the Turkish Academy of Sciences (T\"UBA) through a Principal Membership Grant.
\vspace{12pt}

\noindent{\bf Conflicts of Interest:} The author declares no conflict of interest.

\ed

For example consider the operator $H(t):\C^2\to\C^2$ given by $H(t)(\xi_1,\xi_2):=\varepsilon(\xi_1,if(t)\xi)$ where $\varepsilon$ is a positive real parameter, and $f:\R\to\R$ is a smooth function. In the standard representation of $\C^2$, $H(t)$ is represented by
    \[\bH(t)=\varepsilon\left[\begin{array}{cc}
    1 & 0\\
    0 & i f(t)\end{array}\right].\]
Because the spectrum of $H(t)$ consists of a real and an imaginary complex number, namely $\varepsilon$ and $i \varepsilon f(t)$, $H(t)$ is not a pseudo-Hermitian operator except for times at which $f(t)=0$. Because $\bH(t)$ is diagonal, we can easily derive the following formula for the standard matrix representation of the time-evolution operator $U(t,t_0)$.
    \be
    \bU(t,t_0)=\varepsilon\left[\begin{array}{cc}
    e^{-i\varepsilon(t-t_0)} & 0\\
    0 & e^{\varepsilon F(t)}\end{array}\right],
    \label{x29}
    \ee
where $F(t):=\int_{t_0}^t f(t')dt'$.

Let us identify $\bigeta_0$ with the identity operator, so that at $t=t_0$, $\sH_{\eta(t)}$ coincides with $\sH$. Then in view of (\ref{x5}), (\ref{xx-28}), and (\ref{x29}), we can show that the metric operator $\bigeta(t)$ and the Hermitian Hamiltonian $\bigh(t)$ are respectively represented by the matrices:
    \begin{align}
    &\boldsymbol{\bigeta}(t)=\left[\begin{array}{cc}
    1 & 0\\
    0 & e^{-2\varepsilon F(t)}\end{array}\right],
    &&\boldsymbol{\bigh}=\left[\begin{array}{cc}
    \varepsilon & 0\\
    0 & 0\end{array}\right].
    \label{x30}
    \end{align}
The pairs $(\sH_{\eta(t)},H(t))$ and $(\sH,\bigh)$ both describe the same two-level system. If, following the standard practice, we take $\bigh$ as the energy observable of this system in the representation $(\sH,\bigh)$, we find that the energy observable $H_\ssE:=\bigrho^{-1}(t)\bigh\bigrho(t)$ in the representation  $(\sH_{\eta(t)},H(t))$ coincides with $\bigh$! This is not generally true for other observables. For example, consider the spin observables $s_j$ in the representation $(\sH,\bigh)$. Their standard matrix representation has the form $\bs_j:=\bsigma_j/2$, where $\bsigma_j$ are Pauli matrices. In the representation $(\sH_{\eta(t)},H(t))$, these observables are given by the operators $S_j:=\bigrho^{-1}(t)s_j\bigrho(t)$ with the following matrix representations.
    \begin{align}
    &\bS_1=\frac{1}{2}\left[\begin{array}{cc}
    0 & e^{-\varepsilon F(t)}\\
    e^{\varepsilon F(t)} & 0\end{array}\right],
    &&\bS_2=\frac{1}{2}\left[\begin{array}{cc}
    0 & -i e^{-\varepsilon F(t)}\\
    i e^{\varepsilon F(t)} & 0\end{array}\right],
    &&\bS_2=\frac{1}{2}\left[\begin{array}{cc}
    1 & 0\\
    0 & -1\end{array}\right]=\bs_3.\nn
    \end{align}